\documentclass[aps,prd,twocolumn,nofootinbib,longbibliography,amsfonts,amssymb,amsmath,superscriptaddress]{revtex4-2}
\usepackage[utf8]{inputenc}
\usepackage{physics}
\usepackage{graphicx}
\usepackage[dvipsnames]{xcolor}
\usepackage{microtype}
\usepackage{url}
\usepackage[caption=false]{subfig}
\usepackage[pdftex]{hyperref}
 \usepackage{soul}
\usepackage[T1]{fontenc}
\usepackage{tikz}
\usetikzlibrary{positioning}
\usetikzlibrary{calc}
\usepackage{diagbox}
\usetikzlibrary{decorations.pathreplacing,calligraphy}

\usepackage{mathtools} 

\usepackage{multirow}

\newcommand{\data}{\mathcal{D}}

\newcommand{\thetaps}{\hat\theta}

\newcommand{\thetapsC}{\hat\theta_{C}}

\newcommand{\thetapsK}{\hat\theta_{K}}
\newcommand{\thetapski}{\hat\theta_{{ik}}}

\newcommand{\setdata}{\{\data_k\}_{k\in K}}
\newcommand{\setdataK}{\data_K}
\newcommand{\setdataKnu}{\data_{K,\nu}}
\newcommand{\setdataC}{\data_C}
\newcommand{\setdataKi}{\data_{K_i}}

\newcommand{\metap}{hyperparameter}

\newcommand{\metaps}{hyperparameters}
\newcommand{\Metaps}{Hyperparameters}

\newcommand{\npost}{n_{\rm post}}

\newcommand{\catalog}{\{d_k\}_{k\in C}}

\newcommand{\qinit}{q_{\rm init}}
\newcommand{\nprop}{N_{\rm prop}}

\newcommand{\trainingsetonenumberPL}{6.7\times10^5}
\newcommand{\trainingsettwonumberPL}{4.4\times10^4}
\newcommand{\resultsonenevents}{60}
\newcommand{\resultstwonevents}{600}

\newcommand{\yvb}{\vb{Y}}
\newcommand{\zvb}{\vb{Z}}
\newcommand{\gvb}{\vb{g}}

\newcommand{\svb}{\vb{s}}

\newcommand{\loss}{\mathbb{L}}

\newcommand{\nbatch}{n_{\rm b}}
\newcommand{\nbatchevents}{n_{\rm sub}}
\newcommand{\nblock}{n_{\rm block}}

\newcommand{\lowevents}{\texttt{low}}
\newcommand{\highevents}{\texttt{high}}

\newcommand{\datasetsample}{hyper-sample}
\newcommand{\datasetsamples}{\datasetsample s}

\newcommand{\natunit}{\rm nat}

\newcommand{\msone}{m_{1,s}}
\newcommand{\mstwo}{m_{2,s}}
\newcommand{\mdone}{m_{1,d}}
\newcommand{\mdtwo}{m_{2,d}}

\newcommand{\ms}{m_{s}}
\newcommand{\md}{m_{d}}
\newcommand{\mmin}{m_\mathrm{min}}
\newcommand{\mmax}{m_\mathrm{max}}

\newcommand{\nobs}{{N_\mathrm{obs}}}

\newcommand{\pdet}{p_\mathrm{det}(\theta)}
\newcommand{\ppop}{{p_\mathrm{pop}(\theta|\Lambda)}}
\newcommand{\ppopK}{p_{\rm pop}(\theta_K|\Lambda)}

\newcommand{\plmassname}{\textsc{Power Law}}

\newcommand{\dint}{\mathrm{d}}

\newcommand{\dingo}{\texttt{dingo}}

\newcommand{\bilby}{\texttt{bilby}}
\newcommand{\icarogw}{\texttt{icarogw}}
\newcommand{\lalinference}{\textsc{LALInference}}

\newcommand{\Omegam}{\Omega_{{\rm m}}}
\newcommand{\msun}{\text{M}_\odot}
\newcommand{\hu}{\,{\rm km \,s^{-1} \, Mpc^{-1}}} %
\newcommand{\phXPHM}{\textsc{IMRPhenomXPHM}}

\newcommand{\newcom}[1]{#1}

\newcommand{\deffrom}{\coloneqq}

\newcommand{\specialcell}[2][c]{%
  \begin{tabular}[#1]{@{}c@{}}#2\end{tabular}}

\newcommand{\specialcellleft}[2][c]{%
  \begin{tabular}[#1]{@{}l@{}}#2\end{tabular}}

\usepackage[acronym, toc, nonumberlist]{glossaries}

\glsdisablehyper
\setacronymstyle{long-short}

\newacronym{ns}{NS}{neutron star}
\newacronym{bh}{BH}{black hole}
\newacronym{bbh}{BBH}{binary black hole}
\newacronym{bns}{BNS}{binary neutron star}
\newacronym{nsbh}{NSBH}{neutron star black hole}

\newacronym{eos}{EoS}{equation of state}
\newacronym{gw}{GW}{gravitational wave}
\newacronym{gr}{GR}{general relativity}
\newacronym{snr}{SNR}{signal-to-noise ratio}

\newacronym{lisa}{LISA}{Laser Interferometer Space Antenna }
\newacronym{ligo}{LIGO}{Laser Interferometer Gravitational wave Observatory}
\newacronym{kagra}{KAGRA}{KAmioka GRavitational wave detector}
\newacronym{eob}{EOB}{effective one-body}
\newacronym{em}{EM}{electromagnetic}
\newacronym{lcdm}{$\Lambda$CDM}{$\Lambda$ cold dark matter}
\newacronym{pl}{PL}{power law}
\newacronym{plg}{PLG}{power law and Gaussian}
\newacronym{kde}{KDE}{kernel density estimate}
\newacronym{de}{DE}{dark energy}
\newacronym{cdf}{CDF}{cumulative density function}

\newacronym{lvk}{LVK}{LIGO-Virgo-KAGRA}
\newacronym{ego}{EGO}{European gravitational observatory}
\newacronym{asd}{ASD}{amplitude spectral density}
\newacronym{psd}{PSD}{power spectral density}
\newacronym{mcmc}{MCMC}{Monte Carlo Markov chain}
\newacronym{hlv}{HLV}{Hanford Livingston Virgo}
\newacronym{pe}{PE}{parameter estimation}
\newacronym{cbc}{CBC}{compact binary coalescence}
\newacronym{aligo}{aLIGO}{advanced LIGO}
\newacronym{far}{FAR}{false alarm rate}

\newacronym{cl}{CL}{confidence level}

\newacronym{pn}{PN}{post-Newtonian}
\newacronym{nr}{NR}{numerical relativity}
\newacronym{ppisn}{PPISN}{pulsation pair-instability supernova}
\newacronym{pisn}{PISN}{pair instability-supernova}
\newacronym{et}{ET}{Einstein Telescope}
\newacronym{ce}{CE}{Cosmic Explorer}

\newacronym{cmb}{CMB}{cosmic microwave background}
\newacronym{lss}{LSS}{large scale structure}
\newacronym{isco}{ISCO}{innermost stable orbit}

\newacronym{oi}{Oi}{observation run $i$}
\newacronym{gwtci}{GWTC-i}{gravitational wave transient catalog $i$}

\newacronym{2g}{2G}{second generation}
\newacronym{3g}{3G}{third generation}

\newacronym{bao}{BAO}{baryonic acoustic oscillation}

\newacronym{wkb}{WKB}{Wentzel–Kramers–Brillouin}

\newacronym{dpg}{DPG}{Dvali-Gabadadze-Porrati}
\newacronym{dhost}{DHOST}{degenerate higher-order scalar-tensor}
\newacronym{mg}{MG}{modified gravity}
\newacronym{des}{DES}{dark energy survey}

\newacronym{tt}{TT}{transverse-traceless}

\newacronym{sgwb}{SGWB}{stochastic gravitational wave background}
\newacronym{dgrb}{DGRB}{diffuse $\gamma$-ray background}
\newacronym{gut}{GUT}{grand unified theory}

\newacronym{ng}{NG}{Nambu-Goto}
\newacronym{gbr}{GBR}{gravitational backreaction}

\newacronym{nf}{NF}{normalizing flow}
\newacronym{ml}{ML}{machine learning}
\newacronym{lfi}{LFI}{likelihood-free inference}
\newacronym{nn}{NN}{neural network}
\newacronym{dingo}{DINGO}{deep inference for gravitational wave observations}
\newacronym{gpu}{GPU}{graphics processing unit}
\newacronym{hba}{HBA}{hierarchical Bayesian analysis}

\newacronym{kl}{KL}{Kullback-Leibler}
\newacronym{js}{JS}{Jensen-Shannon}
\newacronym{ks}{KS}{Kolmogorov–Smirnov}

\newacronym{smbh}{SMBH}{supermassive black hole}
\newacronym{agn}{AGN}{active galactic nuclei}
\newacronym{spa}{SPA}{stationary phase approximation}

\newacronym{pta}{PTA}{pulsar timing array}

\newacronym{npe}{NPE}{neural posterior estimation}

\usepackage{aas_macros} 
\usepackage[strings]{underscore}

\begin{document}

\title{Gravitational wave populations and cosmology with neural posterior estimation}

\author{Konstantin Leyde}
\email{konstantin.leyde@port.ac.uk}
\affiliation{Universit\'e Paris Cit\'e, CNRS, Astroparticule et Cosmologie, F-75013 Paris, France}
\affiliation{Institute of Cosmology and Gravitation, University of Portsmouth, \\
Burnaby Road, Portsmouth PO1 3FX, United Kingdom}
\author{Stephen R. Green}
\email{stephen.green2@nottingham.ac.uk}
\affiliation{School of Mathematical Sciences, University of Nottingham\\
  University Park, Nottingham NG7 2RD, United Kingdom}
\author{Alexandre Toubiana}
\email{alexandre.toubiana@aei.mpg.de}
\affiliation{Max Planck Institute for Gravitational Physics (Albert Einstein Institute)\\
  Am M\"uhlenberg 1, 14476 Potsdam, Germany}
\author{Jonathan Gair}
\email{jonathan.gair@aei.mpg.de}
\affiliation{Max Planck Institute for Gravitational Physics (Albert Einstein Institute)\\
  Am M\"uhlenberg 1, 14476 Potsdam, Germany}

\begin{abstract}
  We apply neural posterior estimation for fast-and-accurate hierarchical Bayesian inference of gravitational wave populations. We use a normalizing flow to estimate directly the population hyper-parameters from a collection of individual source observations. This approach provides complete freedom in event representation, automatic inclusion of selection effects, and (in contrast to likelihood estimation) without the need for stochastic samplers to obtain posterior samples. Since the number of events may be unknown when the network is trained, we split into sub-population analyses that we later recombine; this allows for fast sequential analyses as additional events are observed. We demonstrate our method on a toy problem of dark siren cosmology, and show that inference takes just a few minutes and scales to $\sim 600$ events before performance degrades. We argue that neural posterior estimation therefore represents a promising avenue for population inference with large numbers of events.

\end{abstract}

\maketitle


\section{Introduction}

Hierarchical Bayesian analysis (HBA) provides the statistical framework to combine individual gravitational wave (GW) observations to answer questions about entire populations. Starting from a population model $p_\text{pop}(\theta|\Lambda)$ for source parameters $\theta$ depending on population \emph{hyper-parameters} $\Lambda$, with prior $p(\Lambda)$, HBA characterizes the population in terms of the posterior distribution $p(\Lambda|\mathcal D_C)$, where $\mathcal D_C$ is a catalog of GW observations. With over 100 observations by the LIGO-Virgo-KAGRA Collaboration \cite{det1-aligo2015, det2-aLIGO:2020wna, det3-Tse:2019wcy, det4-VIRGO:2014yos, det5-Virgo:2019juy, det6-KAGRA:2020tym, det7-Aso:2013eba} to date~\cite{LIGOScientific:2021djp}, HBA has been used to constrain a wide variety of population properties including mass and spin distributions~\cite{Fishbach:2017dwv, LIGOScientific:2018jsj, Isi:2019asy, LIGOScientific:2020kqk, Romero-Shaw:2020thy, Romero-Shaw:2020aaj, Tiwari:2020otp, KAGRA:2021duu, Edelman:2021zkw, Hoy:2021rfv, Biscoveanu:2022qac, Ashton:2021tvz, Bavera:2022mef, Karathanasis:2022rtr, Fishbach:2022lzq, Mastrogiovanni:2022ykr, Ye:2022qoe, Fishbach:2022lzq, Antonini:2022vib, Adamcewicz:2022hce, Farah:2023vsc, Callister:2023tgi, Fishbach:2023xws, Mould:2023ift, Sadiq:2023zee, Rinaldi:2023bbd}, and fundamental physics \cite{LIGOScientific:2019fpa, LIGOScientific:2020tif, LIGOScientific:2021sio, MaganaHernandez:2021zyc, Ezquiaga:2021ayr, Mancarella:2021ecn, Leyde:2022orh, Isi:2022cii, Niu:2022yhr, Chen:2023wpj, Ray:2023sbr, Mastrogiovanni:2023emh, Magee:2023muf, Payne:2023kwj}.

When combined with redshift information, GWs can also be used to constrain cosmology. Indeed, the joint GW and electromagnetic observation of GW170817---a standard \emph{siren}---constrained the Hubble constant $H_0$ to within $\sim 20\%$~\cite{LIGOScientific:2017vwq,LIGOScientific:2017adf}. However, the vast majority of observations are of binary black holes, with no electromagnetic counterpart. In these cases, statistical \emph{dark siren} methods using HBA can nevertheless still place constraints on cosmology. This can be done by either correlating GW signals with galaxy catalogs~\cite{Schutz:1986gp, DelPozzo:2011vcw, Gray:2019ksv, Gray:2021sew, Finke:2021aom, Turski:2023lxq, Mastrogiovanni:2023emh, Gray:2023wgj} or by involving assumptions on the source binary mass distribution~\cite{Taylor:2011fs, Taylor:2012db, Farr:2019twy, Mastrogiovanni:2021wsd, Mancarella:2021ecn, Leyde:2022orh, Ezquiaga:2022zkx}.

Here, we focus on the mass spectrum method: given a population model in the source frame, the predicted distribution of detector frame masses\footnote{Recall the relation between detector frame masses $\md$ and source frame masses $\ms$ are related as $\md = (1+z)\ms$. Throughout, we assume the contribution from proper motion to be negligible against the cosmological redshift.} and luminosity distance depends on the population and the cosmological parameters. By comparing this predicted distribution to the one observed with \glspl{gw}, we can therefore jointly constrain population and cosmological parameters.

However, the current uncertainty on $H_0$ from GW observations is much larger than from studies of the cosmic microwave background \cite{Planck:2018vyg} or supernovae \cite{Riess:2019cxk} and it will not be before $\mathcal{O}(10^4)$ binary black hole mergers \cite{Farr:2019twy,Mastrogiovanni:2021wsd,Ezquiaga:2022zkx,Mancarella:2021ecn}, or several hundred binary neutron stars \cite{Chen:2017rfc} that their uncertainty will be comparable.
Networks of future detectors, such as the Einstein telescope (ET) and Cosmic Explorer (CE), will provide the requisite large number of observed events, reaching far into the cosmic past. This will allow for the precise inference of cosmological parameters, using bright sirens \cite{Belgacem:2019tbw}, in conjunction with galaxy catalogs \cite{Yu:2020vyy, Borhanian:2020vyr, Song:2022siz, Gupta:2022fwd, Zhu:2023jti, Muttoni:2023prw} and features in the mass spectrum \cite{Taylor:2011fs, Taylor:2012db, You:2020wju, Leandro:2021qlc, Muttoni:2023prw}. 
Conventional population analyses (hierarchical Bayesian inference methods) require an analytic population model, and are slow when analyzing a large number of events. 
The large number of events of the upcoming detector networks calls for new methods for the measurement of the \metaps{} (e.g.,~$H_0$) with \gls{gw} events.\footnote{Since the \metaps{} describe the overall distribution of source parameters rather than the single-event ones, the extraction of the \metaps{} is also referred to as \textit{hierarchical inference}.}

In this work, we apply neural posterior estimation to population inference of \gls{gw} signals. 
The specific illustrative problem we set out to solve is to obtain constraints on cosmological parameters through the dark siren mass spectrum method\footnote{We note, however, that the proposed method of population analysis with deep neural networks is not limited to this application. In principle, our scheme could use electromagnetic, or \gls{gw} data or both to produce constraints on the cosmological or population parameters.}, addressing the aforementioned issues. 
In addition to the gain in computational speed,\footnote{\newcom{Note that just-in-time compilation, the use of GPUs and gradient-based sampling algorithms can achieve similar speeds.}} simulation-based approaches can, in principle, directly incorporate predictions from astrophysical simulations, without having to resort to phenomenological descriptions of the resulting source parameter distributions.
We summarize the analyzed \gls{gw} data by posterior samples of the parameters of the individual events\footnote{In the following, we refer to these as single-event posterior samples. }, but the method could be applied to any input data that summarizes the GW observations sufficiently well. 
It is therefore particularly adapted to future analysis chains that rely on other deep learning algorithms. 
In principle, our method can also account for additional uncertainty from latent variables, which are difficult to account for in conventional methods or when modeling the population likelihood.
For example, this could include the use of different waveforms for the production of single-event posterior samples.


The learning task is to approximate the posterior $p(\Lambda|\setdataC)$, where $\Lambda$ is the set of \metaps{} describing the population model and the cosmological parameters, and $\setdataC$ is the GW catalog data.
We propose a deep neural network scheme that learns directly the posterior distribution of the population parameters -- including the selection effect. 
In particular, this approach allows us to infer population properties in a likelihood-free way (also referred to as \textit{simulation-based inference}), requiring solely the simulation of observed event data.


A number of previous studies have applied machine learning techniques to aspects of the population inference problem. In~\cite{Talbot:2020oeu}, machine learning was used to estimate the selection function, while \cite{Wong:2020jdt, Gerardi:2021gvk, Mould:2022ccw, Ruhe:2022ddi} used machine learning to represent the \textit{population likelihood} (including selection effects in the latter two cases). By contrast, in our approach we directly model the \textit{population posterior distribution}, which circumvents the need for an additional MCMC analysis to obtain the \metap{} posterior since posterior samples are produced directly through importance sampling. 
Additionally, \cite{Gerardi:2021gvk} learned the population likelihood (in the bright siren case -- assuming an \acrshort{em} counterpart), but used a toy model for the single-event posterior distribution, whereas our method uses posterior samples generated with the realistic deep learning model \dingo{} \cite{Dax:2021myb}. It has been shown that this model agrees very well with the true posterior distribution in the parameter ranges we consider.

The network's architecture used here is that of a \textit{conditional normalizing flow} \cite{tabak2010density, tabak2013family, dinh2014normalizingflows, 2015arXiv150505770J, KobyzevPAMI2020}. This framework allows one to generate a distribution conditioned on data, and to draw samples from the distribution efficiently. This method has been applied to a large variety of problems in science \cite{Kanwar:2020xzo,scienceaaw1147,wirnsberger2020targeted,kohler2019equivariant, Wong:2020jdt,Green:2020hst, Green:2020dnx, Dax:2021myb, Dax:2021tsq}. In particular, it has accelerated single-event parameter estimation for compact binary coalescences by several orders of magnitude, see \dingo{} \cite{Green:2020hst,Green:2020dnx, Dax:2021tsq,Dax:2021myb}. 
Whereas the latter model outputs posterior samples of individual event parameters given the estimated noise spectral density and measured strain for that event, the model described in this work outputs the distribution of the \metaps{} given posterior samples of the individual events. 

The structure of this paper is as follows. In Section~\ref{sec: method}, we begin by revisiting the classical approach to population inference utilizing Bayesian statistics. Following this, in Section~\ref{subsec: divide and conquer}, we outline our divide-and-conquer strategy, which splits the population into smaller sub-populations for independent analysis, subsequently merging them to obtain the final result analysing the complete catalog. In Section~\ref{sec: astrophysical setup}, we then provide an overview of the astrophysical assumptions that underlie our study. The training dataset, along with its number of entries, is then presented in detail in Sec.~\ref{sec: training sets}. From these training datasets, we train our models and present the results, which are described in Section~\ref{sec: results}, accompanied by a comparison against the traditional Bayesian approach. Finally, in Section~\ref{sec: conclusion}, we discuss our results and possible extensions to our work.

\begin{table}[]
    \centering
    \renewcommand{\arraystretch}{1.4}
    \begin{tabular}{cl}
  \hline
  \hline
  \textbf{Variable} & \textbf{Description} \\
  \hline
  \multicolumn{2}{c}{\textbf{GW data}}\\
  \hline
  $\theta$ & Single-event BBH parameters \\
    $p(\thetaps|\data)$ & Single-event posterior distribution \\
    \hline
    \multicolumn{2}{c}{\textbf{GW catalog}}\\
  \hline
    $\data_C$ & A catalog $C$ of GW observations \\
    $\nobs$ & Number of observed GW events \\
    \hline
    \multicolumn{2}{c}{\textbf{Population parameters}}\\
  \hline
  $\Lambda$ & \Metaps{} \\
  $\ppop$  & Population model \\
  $p(\Lambda|\data_C)$  & \specialcellleft{Hyperparameter posterior from\\a catalog $C$, cf.~Sec.~\ref{eq: population posterior}} \\
  $\xi(\Lambda)$  & Selection bias \\
  \hline
  \multicolumn{2}{c}{\textbf{Machine learning}}\\
  \hline
  $\nbatchevents$ & Number of events per sub-population \\
  $q(\Lambda|\data_C)$ &
  \specialcellleft{\Metaps{} posterior estimate\\from a GW catalog $C$} \\
  \hline
  \hline
\end{tabular}
    \caption{Overview of the variables and quantities used. }
    \label{preface: tab: overview variables}
\end{table}

\section{Methods}

We now outline the conventional hierarchical Bayesian population analysis and relate it to the deep neural network approach in Sec.~\ref{subsec: npe methods}. 
The classical approach will function as our reference point against which we will compare the outcomes with the \gls{nf} method.
We refer to the classical method as HBA (hierarchical Bayesian analysis) and to the neural network model as \gls{npe}.

To facilitate the following discussion, we introduce some notation (see also tab.~\ref{preface: tab: overview variables} for a summary of the variables used). 
We denote the set of \metaps{} as $\Lambda$ -- this can include cosmological parameters such as $H_0$ and the parameters describing the mass, spin and redshift distribution of individual events. 
The true source parameters are written as $\theta$, and the distribution of data, $\data$, given the true parameters as $p(\data|\theta)$. 
The latter term is the single-event \gls{gw} likelihood. The population model is denoted as $\ppop$, and we use $K$ to denote a collection of events. For instance, $\theta_K\deffrom \{\theta_i\}_{i\in K}$ is the set of true parameters of events in the set $K$. 
In this notation, the probability of drawing the true parameters $\theta_K$ from the population model is then
\begin{equation}
    \ppopK = \prod_{j\in K}p_{\rm pop}(\theta_j|\Lambda)\,,
\end{equation}
since individual sample draws are independent. 
%
The number of events in the \gls{gw} catalog is denoted as $\nobs\deffrom {\rm card}(C)$,  with ${\rm card}(X)$ the number of elements in the set $X$. 

\subsection{Hierarchical Bayesian population method}
\label{sec: method}

The goal of extracting population and cosmological parameters from \gls{gw} data is classically approached with a \gls{hba} \cite{LIGOScientific:2018jsj, LIGOScientific:2020kqk, KAGRA:2021duu, LIGOScientific:2021aug, Farr:2019twy, Ezquiaga:2018btd, Ezquiaga:2020tns, Mastrogiovanni:2021wsd, 2021arXiv210908748C, 2021arXiv210405139M, 2021arXiv211207650M}.
We wish to infer the posterior distribution of $\Lambda$, based on a set of GW events $\setdataC \deffrom \{\data_i\}_{i\in C}$. 
With the catalog $\setdataC$ the posterior of $\Lambda$ can be rewritten with Bayes' theorem as $p(\Lambda|\setdataC)= p(\Lambda)p(\setdataC|\Lambda) /p(\setdataC) $, where $p(\Lambda)$ denotes the prior knowledge of $\Lambda$, $p(\setdataC|\Lambda)$ is the hierarchical likelihood, and $p(\setdataC)$ is the evidence, the probability of observing data $\setdataC$.

Using then the \gls{hba} scheme, the posterior of the \metaps{} informed from $\nobs$ events is given by (marginalizing over the overall rate of events) \cite{Mandel:2018mve,2019PASA...36...10T,Vitale:2020aaz}
\begin{align}
\label{eq: population posterior}
    p(\Lambda|\setdataC) &= \frac{p(\Lambda)}{p(\setdataC)} 
    p(\setdataC|\Lambda) = 
    \frac{p(\Lambda)}{p(\setdataC)} 
    \prod_{j=1}^{\nobs} 
    p(\data_j|\Lambda)
    \nonumber
    \\
    &=\frac{p(\Lambda)}{p(\setdataC)}
    \prod_{j=1}^{\nobs} \frac{\int p(\data_j|\theta_j)\,p_\mathrm{pop}(\theta_j| \Lambda) \mathrm{d}\theta_j }{\int p_\mathrm{det}(\theta_j)\,p_\mathrm{pop}(\theta_j| \Lambda) \mathrm{d}\theta_j }\,,
\end{align}
the prior on $\Lambda$ is denoted as $p(\Lambda)$ and the prior probability of the data as $p(\setdataC)$. 
The uncertainty in our knowledge of single-event parameters is encoded in the likelihood $p(\data_j|\theta_j)$ of obtaining data $\data_j$, given the true parameters $\theta_j$. 
Finally, the probability of detection, given the source parameters $\theta$, is denoted by $\pdet$ and depends (amongst other factors) on the detector sensitivity, the number of detectors and the detection threshold. This encodes the fact that not all data is included in the set $\setdataC$, but we choose segments of data in which we are confident that signals of astrophysical origin are present. This selection is a property of the data alone. The data is the sum $\data = h(\theta)+n$ of the pure signal $h(\theta)$ and the noise $n$. The detection probability is the probability that this data lies in the region we define as a detected source,  
i.e., $\pdet=\int_{\data\text{ detected}}\mathrm{d}n\,p(\mathcal{D}|\theta)$. 
The denominator (in the product) of the above equation accounts for this \textit{selection effect} -- not all GW sources have the same probability of detection. It is common to define the detected fraction of the population $\xi(\Lambda)\deffrom\int p_\mathrm{det}(\theta)\,p_\mathrm{pop}(\theta| \Lambda) \mathrm{d}\theta$. 
In general, it is difficult to evaluate this term, and one usually relies on an injection campaign to produce a set of detected \gls{gw} signals. 
We will show that our method accounts for the selection effect, bypassing the explicit computation of $\xi(\Lambda)$. Effectively, we perform an injection campaign during the generation of the training data and hence, the cost is amortized over the repeated evaluation of the neural population posterior. 

There is some freedom in the representation of the \gls{gw} data $\data_i$: 
we focus here on posterior samples, that approximate the uncertainty of the source parameters $\theta$ (such as the component source frame masses, or the luminosity distance). 
The posterior samples follow the distribution $\thetaps \sim p(\thetaps|\data_k)$, and we denote the assumed prior under which the posterior samples were created as $\pi_{\rm MCMC}$. 
In the following, we use $\thetapski$ to denote the $i$th posterior sample from the GW event $k$ (compare to Eq.~\eqref{eq: def posterior samples theta notation}), and $n_{{\rm post},k}$ is the number of posterior samples for this event. 
The numerator of Eq.\,\eqref{eq: population posterior} is usually approximated by summing over posterior samples of the individual GW events.
The population likelihood as informed by \textit{one} GW event $\data_k$ can then be rewritten as
\begin{equation}
\label{eq: indivual population likelihood}
    p(\data_k|\Lambda) 
     \approx \frac{p(\data_k)}{\xi(\Lambda)\,n_{{\rm post},k}}\sum_{i=1}^{n_{{\rm post},k}}\frac{p_{\rm pop}\left(\thetapski|\Lambda\right)}{\pi_{\rm MCMC}\left(\thetapski\right)}\,,
\end{equation}
where the sum above is taken over the posterior samples $\thetapski\sim p(\thetapski|\data_k)$. 
To evaluate the full population posterior of Eq.\,\eqref{eq: population posterior}, one multiplies the individual contributions of Eq.\,\eqref{eq: indivual population likelihood}.


\subsection{Neural posterior estimation (NPE) methods} 

\label{subsec: npe methods}

Hierarchical Bayesian analysis becomes increasingly expensive as the number of sources included in the analysis increases, due both to the cost of obtaining the posterior samples for each event, and the cost of combining the events to obtain the population posterior. The use of machine learning approaches is becoming increasingly widespread in the physical sciences, as these often provide a fast and efficient way to complete complex analysis tasks. In a gravitational wave context, \dingo~ has been shown to generate posterior distributions nearly indistinguishable from those produced by standard sampling algorithms in a small fraction of the time~\cite{Dax:2021tsq}, while residual differences can be efficiently eliminated through importance sampling~\cite{Dax:2022pxd}. We hope to see similar benefits from the application of machine learning methods to population inference. A major complication is that the number of events that will be observed is not typically known a priori. Not only does this present the difficulty of generating an arbitrarily large training dataset, but neural networks typically have fixed input dimension. We overcome this problem by implementing a strategy that divides the GW catalog into smaller sub-populations, each containing $\mathcal{O}(10-100)$ signals.
Our model then learns the posterior distribution \textit{analyzing a sub-population of events}. We combine the intermediate results (the population posterior of each sub-population) to derive the population posterior of the entire catalog.\footnote{The \metaps{} samples are combined via importance sampling, as detailed in Sec.~\ref{subsec: divide and conquer}. }
%
We will now elaborate on the model loss, how to combine sub-populations of events, the \gls{nf}'s architecture and the generation of the training dataset.

\subsubsection{Sub-population analysis}
\label{subsec: divide and conquer}

To simplify the problem, we split the \gls{gw} catalog into smaller sub-populations. 
Calling one of these sub-populations $\setdataK\deffrom\setdata$,
the model we propose then approximates the population posterior from analyzing $\setdataK$, converging to the term $p(\Lambda|\setdata)$. 
One then obtains the complete posterior analyzing all events by combining the individual posteriors of each of the sub-populations.
This approach ensures the computational cost to generate the training dataset is not too large. 


The catalog $C$ is divided into sub-populations of events, $\left\{ K_i\right\}$, where each of the $K_i$ contains $\nbatchevents$ events.\footnote{Throughout, we assume the length of the subset of events $\nbatchevents$ to divide the total number of events $\nobs$.} 
That is, the $K_i$, for $i\in\{1,2,\ldots, \nbatch\}$, define a (random) distinct partition of $C$, i.e.
\begin{equation}
    C = 
    K_1 \,\dot{\cup} \,K_2 \,\dot{\cup} 
    \; \ldots \; 
    \dot{\cup} \,K_{\nbatch}\,,
\end{equation}
with $\nbatch\deffrom \nobs/\nbatchevents$. 
The machine learning model produces a population posterior $q(\Lambda|\setdataKi)$ for each of the sub-populations, which approximates $p(\Lambda|\setdataKi)$.
The repeated application of Bayes' theorem yields the complete posterior informed by all events in $C$, i.e.
\begin{equation}
\label{methods: machine learning: eq: approximation full posterior}
    q(\Lambda|\setdataC) \deffrom \frac{\mathcal{N}}{p(\Lambda)^{\nbatch-1}}\prod_{i=1} ^ {\nbatch}q(\Lambda|\setdataKi)\,,
\end{equation}
with $p(\Lambda)$ the prior on the \metaps{} and $\mathcal{N}$ is a normalization constant given by $\mathcal{N}^{-1}\deffrom \int \left[ \prod_{i=1} ^ {\nbatch}q(\Lambda|\setdataKi)/p(\Lambda)^{\nbatch-1} \right]\; {\rm d}\Lambda$. In the limit $q(\Lambda|\setdataC) \deffrom p(\Lambda|\setdataC)$, $\mathcal{N}=\left(\prod_{i=1} ^ {\nbatch}p(\setdataKi)\right)/
p(\setdataC)$, where 
\begin{align}
    p(\setdataKi) &= \int p(\setdataKi|\Lambda) p(\Lambda) \; {\rm d}\Lambda \nonumber \\
    p(\setdataC) &= \int \left[ \prod_{i=1}^{\nbatch} p(\setdataKi|\Lambda) \right] p(\Lambda) \; {\rm d}\Lambda \,.
\end{align}
Below in Sec.~\ref{sec: training sets}, we assume a uniform prior of the \metaps{} $\Lambda$ so that the denominator in equation~(\ref{methods: machine learning: eq: approximation full posterior}) also amounts to a normalization constant.
If the model correctly learns the posterior distribution that analyzes a sub-population of events, we should have the approximation 
\begin{equation}
    q(\Lambda|\setdataC) 
    \approx 
    p(\Lambda|\setdataC)\,.
\end{equation}
The target distribution is conditioned on the observed data. In general, this could be a large space, making the learning task more complex. However, not all components of the data are informative about the target distribution. It is clear from the form of a standard \gls{hba}, Eq.~(\ref{eq: population posterior}), that one possible summary of the data for each event is the set of samples from the individual event parameter posterior distribution. Therefore, 
we make the choice to represent the input data via a set of posterior samples for the \gls{gw} events. 
The \gls{nn} then learns the population posterior from the posterior samples of the individual signals in one sub-population. 
We denote the set of posterior samples of the events in $K$ as $\thetapsK$ and the number of posterior samples per event as $\npost$, assumed to be equal for all events. 
We define, 
\begin{equation}
    \thetapsK = \left\{ \hat\theta_{ij}: \; i\in K; \; j=1,2,\ldots ,\npost-1,\npost\right\}\,,
\end{equation}
where
\begin{equation}
\label{eq: def posterior samples theta notation}
    \hat\theta_{i\bullet} \sim p(\theta|\data_i)\,,
\end{equation}
for $i$ an event in the sub-population $K$. 
From our choice of the data representation, we can then schematically write 
\begin{equation}
    q(\Lambda|\setdataK) 
    \approx
    q(\Lambda|\thetapsK)\,.
\end{equation}
In principle, however, the network could learn the population posterior from any representation of the data $\setdataK$ that is sufficiently informative; this could be the Fourier-transformed or the time-domain strain data. Of course, no matter the representation of the data, the resulting posterior distribution should be the same.

The neural networks used in this work have $\mathcal{O}(10^{6-8})$ parameters that are optimized during the training process to minimize the chosen loss function,  ensuring that the learned function converges to the desired distribution. We take the loss function to be proportional to the \gls{kl} divergence (up to an additive constant), which is defined as
\cite{Kullback:1951zyt}
\begin{equation}
\label{methods: machine learning: eq: def: kullback leibler divergence}
    D_{\rm KL}(p \parallel q) 
    \deffrom
    \int p(x) \log\left(\frac{p(x)}{q(x)}\right)\dint x\,.
\end{equation}
The \gls{kl} divergence is positive semi-definite, and is zero only if $p = q$. Also, note that the \gls{kl} divergence is not symmetric in the distributions $p$ and $q$. Thus, it can be seen as a (generalized) distance between the target distribution and the one learned by the network.

The objective is thus to minimize $D_{\rm KL}(p(\Lambda|\setdataK) \parallel q(\Lambda|\setdataK)) $. 
In reality, we will be approximating $p(\Lambda|\setdataK)$ by $p(\Lambda|\thetapsK)$, since we assume that the data $\setdataK$ is summarized accurately by the single-event posterior samples $\thetapsK$. 
This can be done by minimizing the loss
\begin{align}
\label{methods: machine learning: eq: theoretical loss for model}
\loss \deffrom
    \mathbb{E}_{p(\Lambda)}
    \mathbb{E}_{\ppopK}
    \mathbb{E}_{p(\data_K|\theta_K)}&
    \mathbb{E}_{p(\thetapsK|\data_K)}\left[\right.
    \nonumber
    \\
    &\left.-\log\left(q(\Lambda|\thetapsK)\right)\right]\,, 
\end{align}
where we have introduced the expectation value 
\begin{equation}
    \mathbb{E}_{p(x|y)}\left[f(x,y) \right] \deffrom
    \int \dint x\, p(x|y) \,f(x,y) \,.
\end{equation}
The right hand side of Eq.\,\eqref{methods: machine learning: eq: theoretical loss for model} is the expectation value over four distributions. 
Averaging over noise realizations, $\data_K$, and population draw, $\theta_K$, we can apply Bayes' theorem successively to obtain the equality (see app.~\ref{app: methods: machine learning: loss lemma})
\begin{equation}
\label{methods: machine learning: eq: loss rewritten 2 expectation values}
    \loss = -\mathbb{E}_{p(\thetapsK)}
    \mathbb{E}_{p(\Lambda|\thetapsK)}\log\left[q(\Lambda|\thetapsK)\right]\,.
\end{equation}
From the definition of the \gls{kl} divergence in Eq.~\eqref{methods: machine learning: eq: def: kullback leibler divergence}, we rewrite the above equation as
\begin{align}
    \loss = \mathbb{E}_{p(\thetapsK)}
    &\left[
        D_{\rm KL}(p(\Lambda|\thetapsK) \parallel q(\Lambda|\thetapsK)) 
        \right. 
        \nonumber
        \\
        - & 
        \left.
        D_{\rm KL}(p(\Lambda|\thetapsK) \parallel 1)  
    \right]\,.
\end{align}
Thus, this expression is (up to a constant and the expectation value over $p(\thetapsK)$) the \gls{kl} divergence between the model $q(\Lambda|\thetapsK)$ and the target distribution $p(\Lambda|\thetapsK)$.
Since the \gls{kl} divergence is minimized for $p=q$, it follows that the above loss is also minimized for $p(\Lambda|\thetapsK)=q(\Lambda|\thetapsK)$, and if the network is properly trained, $q(\Lambda|\thetapsK)$ will approximate $p(\Lambda|\thetapsK)$. 
If a network achieved the minimum loss for every possible choice of input parameters, $\thetapsK$, then it would perfectly represent the population posterior. In practice, this will not be achievable. By averaging the loss over noise realizations, $\data_K$, and population draws, $\theta_K$, we ensure that learning effort is expended to represent the distribution best for values of $\thetapsK$ that are more likely to be observed in practice.

To evaluate the loss value of Eq.~\eqref{methods: machine learning: eq: theoretical loss for model} one has to evaluate an expectation value over four distributions. We approximate these expectation values by Monte Carlo averaging, i.e.
\begin{equation}
\label{eq: approximation of expectation values with sums}
    \loss
    \approx
    \frac{1}{N}\sum_{\{\Lambda_\nu,\setdataKnu\}}q(\Lambda_\nu|\setdataKnu)\,,
\end{equation}
where $N$ is the number of samples drawn as follows: according to the prior $p(\Lambda)$ we draw population parameters. For each sample $\Lambda$, we create the cosmological model, draw $\nbatchevents$ true events, simulate $\nbatchevents$ observed strains (passing some specified selection threshold) and produce $\npost$ posterior samples. 
For computational reasons, we precompute the samples $\{\Lambda_\nu,\setdataKnu\}$ and call the resulting data the \textit{training dataset}. 
The loss is then minimized over choices for the \gls{nn} parameters during the training process. 

Note that at no point in the process is the (true) population posterior explicitly evaluated. The above scheme relies solely on the \textit{simulation of data} rather than on evaluating the hierarchical Bayesian likelihood in Eq.~\eqref{eq: population posterior}. 
\newcom{As a consequence, it does not require an analytic expression for the population prior, but solely relies on a forward model to generate training data (i.e.~samples from the population likelihood).
This differs from most HBAs, with the exception of \cite{Wysocki:2020myz,Golomb:2021tll}, which instead use Monte Carlo integration to evaluate the population likelihood, using the population prior to draw samples. In turn, that approach requires to be able to efficiently estimate the single-event likelihood. }

Also, by construction, the model contains the selection effect term $\xi(\Lambda)$ appearing in the denominator of Eq.~\eqref{eq: population posterior}. We thus avoid the computation of this term during inference.\footnote{\newcom{To obtain constraints on a different population distribution requires the training of a new model, which also entails the generation of new training data. This is different from conventional analysis, where the original injection set (to evaluate the selection effect) can be recycled, provided that it covers the parameter space of the new population model sufficiently well. }}

In some cases the \gls{npe} results differ from the \gls{hba} approach for reasons we elaborate below. These differences can be corrected by reweighting the \gls{npe} samples to the target \gls{hba} posterior using importance sampling weights
\begin{equation}
\label{eq: is weights classical recovery}
    w(\Lambda) = \frac{p(\Lambda|\setdataC)}{q(\Lambda|\setdataC)}\,.
\end{equation}
This is possible because we have access to the learned \gls{npe} posterior density, and have an explicit expression for the target \gls{hba} density. 
We show this procedure on one example in Sec.~\ref{subsec: results training set 1}. 
Importance-sampling can also provide a validation: an unchanged posterior (after reweighting) implies that the model has learned the correct \gls{hba} distribution.

\subsubsection{Combining sub-populations of events}
\label{methods: machine learning: subsec: combining events}

In the previous section, we have subdivided the complex problem of obtaining the posterior distribution from catalogs of \gls{gw} events into multiple simpler problems, namely to obtain the posterior distribution from a sub-population of \gls{gw} events. 
We thus train a model $q$ to approximate the population parameter posterior informed by a \textit{subset} of events $\setdataKi$, i.e. $p(\Lambda|\setdataKi)$. One is eventually interested in the posterior as informed by the event catalog $C=\dot \bigcup_{i  = 1}^{\nbatch}K_i$. 
To obtain this distribution we apply the following procedure: 
\begin{enumerate}
    \item With the model, we draw $\nprop$ $\Lambda$ samples from each of the posteriors $q(\Lambda|\setdataKi)$ analyzing a sub-population of \gls{gw} events -- these are our proposal samples. In total, we have $\nbatch \times \nprop$ samples. 
    \item Out of these, we randomly choose $\nprop$ samples. The chosen samples follow the distribution $\qinit(\Lambda|\thetapsC)\deffrom \frac{1}{\nbatch}\sum_{i=1}^{\nbatch} q(\Lambda|\setdataKi)$. 
    \item We evaluate the combined population posterior according to Eq.\,\eqref{methods: machine learning: eq: approximation full posterior} for the proposal samples with our model; to obtain $q(\Lambda|\thetapsC)$. From this, we can compute the weights $w$ as 
    \begin{equation}
    \label{eq: def weights reweighting procedure}
        w(\Lambda|\thetapsC) = \frac{q(\Lambda|\thetapsC)}{\qinit(\Lambda|\thetapsC)} 
        = 
        \frac{\frac{\mathcal{N}}{p(\Lambda)^{\nbatch-1}}\prod_{i=1} ^ {\nbatch}q(\Lambda|\setdataKi)}{\frac{1}{\nbatch}\sum_{i=1}^{\nbatch} q(\Lambda|\setdataKi)}
        \,,
    \end{equation}
    where we applied the definition of $q(\Lambda|\thetapsC)$ in Eq.~\eqref{methods: machine learning: eq: approximation full posterior} in the second equality.
    \item The samples are importance-weighted according to $w(\Lambda|\thetapsC)$ above. The reweighted samples follow the desired distribution $q(\Lambda|\thetapsC)$.
\end{enumerate}
In order to apply this procedure it is vital that one can sample from the distribution and that one has access to the probability with which the samples are created. The architecture of a normalizing flow allows for this. The generation of random samples with normalizing flows is rapid, making the scheme fast. 
We will apply the procedure in practice and compare it to the conventional \gls{hba} method in Sec.~\ref{sec: results}. 

Other scheme are also possible: one could multiply the hierarchical (neural) posterior (dividing out the prior, cf.~Eq.~\ref{methods: machine learning: eq: approximation full posterior}) that analyze each of the sub-populations (the probability of which is given by the flow) and use MCMC sampling to recover the combined posterior (that analyzes the entire catalog).
The sampling method we outlined above avoids running a full MCMC analysis, which would further increase the computing time. We have compared the two approaches for selected cases and found very good agreement.


\subsection{Flow architecture}
\label{subsec: flow architecture}

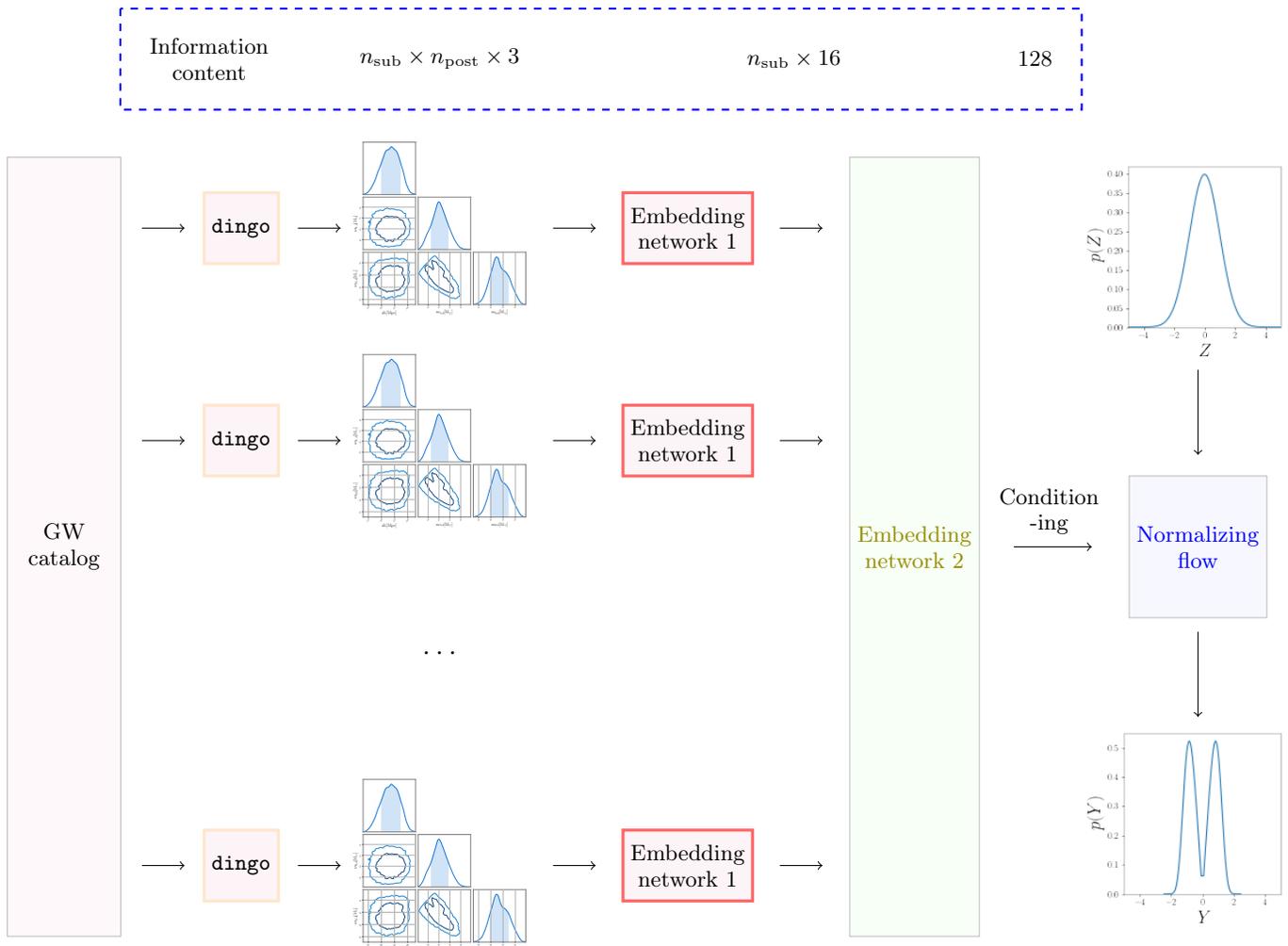
\begin{figure*}[]
    \centering
    \begin{tikzpicture}[
    squarednode/.style={rectangle, draw=black!60, fill=blue!5, very thick, minimum size=10mm, scale = 1},
    pics/rr/.style={code={
    \draw[] (0,0) -- (0.2,0) -- (0.2,0.2) -- (0.0,0.2) -- cycle;
    \draw[->] (-4.2,0) -- (-3.6,0);
    \node[squarednode,align=center, draw=orange!20, fill=magenta!5] at (-2.8,0) {\dingo};
    \draw[->] (-2.,0) -- (-1.4,0);
    \node[inner sep=0pt] (corner) at (0,0) {\includegraphics[width=2.7cm]{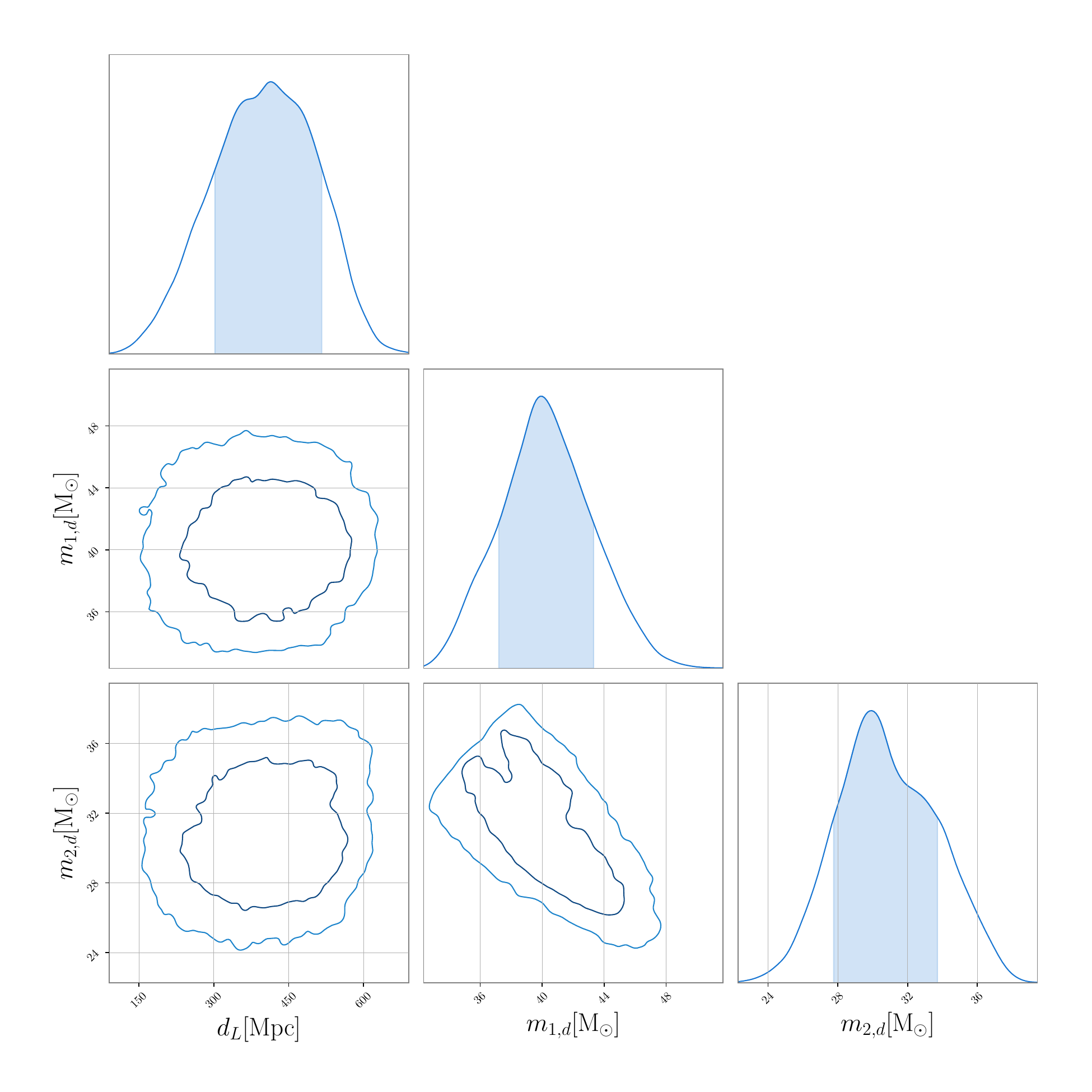}};
    \draw[->] (1.6,0) -- (2.2,0);
    \node[squarednode,align=center, draw=red!60, fill=magenta!5] at (3.5,0) {Embedding\\network 1};
    \draw[->] (4.8,0) -- (5.4,0);
    }},]
    \node[rectangle,
        draw = lightgray,
        text = black,
        fill = magenta!3,
        minimum width = 1.6cm, 
        minimum height = 11cm,
        align = center] (r) at (-11.3, 4.5) {GW\\catalog};

    \path foreach \X in {0,2,3}
 { ($(0,0)+(-6.0, 3 * \X)$) pic{rr}};
    \node[align=center] (punktpunkt) at (-6,3) {\large $\ldots$}; 
    \node[align=center] (input) at (-9.25,7.75) {};

    \node[yshift = 7.cm] (x) [above =of punktpunkt]{$\nbatchevents\times\npost \times 3$};
    \node[align=center] (orderofmagn) at (x-|input) {Information\\content};
    \node[align=center, xshift = 5cm] (embinformation) at (x) {$\nbatchevents\times 16$};
    \node[rectangle,
        draw = lightgray,
        text = olive,
        fill = green!3,
        minimum width = 1cm, 
        minimum height = 11cm,
        align = center] (emb2) at (0.7, 4.5) {Embedding\\network 2};
    \pgfmathsetmacro{\xoffset}{-1.3}
    \node[align=center] (embinformation2) at ($(x-|emb2) + (1.7, 0)$) {$128$};
    \draw[thick,dashed,draw = blue] ($(orderofmagn.north west)+(-0.3,+0.3)$) rectangle ($(embinformation2.south east)+(0.3,-0.5)$);
    \node[align=center] at (3.9+\xoffset,5) {Condition\\-ing};
    \draw[->] (3.4+\xoffset,4.5) -- (4.5+\xoffset,4.5);
    \node[rectangle,
    draw = lightgray,
    text = blue,
    fill = blue!3,
    minimum width = 1cm, 
    minimum height = 2cm,
    align = center] (nflow) at (6+\xoffset, 4.5) {Normalizing\\flow};
    \draw[->] (6+\xoffset,7) -- (6+\xoffset,5.8);
    \draw[->] (6+\xoffset,3.3) -- (6+\xoffset,2.1);
    \coordinate (nflowZplotcoord) at (5.8+\xoffset,8.5);
    \coordinate (nflowYplotcoord) at (5.8+\xoffset,0.5);
    \node[inner sep=0pt] (nflowZplot) at (nflowZplotcoord) {\includegraphics[width=3cm]{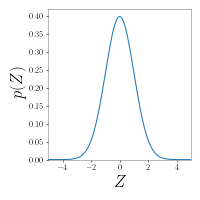}};
    \node[inner sep=0pt] (nflowYplot) at (nflowYplotcoord) {\includegraphics[width=3.cm]{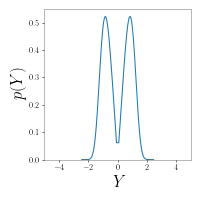}};
    \end{tikzpicture}
    \caption{Overview of the data reduction with the embedding networks and the conditioning of the normalizing flow. This method reduces the data dimension from initially $\nbatchevents\times\npost\times 3$ to $128$. Note that the embedding network 1 is identical for all \gls{gw} events. This data summary reduces the number of adjustable \gls{nn} parameters and hence simplifies the training process. }
    \label{ml: fig: our method and embedding networks}
\end{figure*}

The following section summarizes the building blocks that make up our \gls{npe} model. 
The proposed machine learning model combines two embedding neural networks for data compression and a normalizing flow for population posterior generation as described in Fig.~\ref{ml: fig: our method and embedding networks}. In the following, we refer to the full algorithm simply as the ``model''.

The posterior samples from all events in one sub-population represent a large dataset that we seek to reduce with two embedding networks that summarize (\textit{i}) the individual events in a first stage and (\textit{ii}) the set of all summaries of $\nbatchevents$ events produced by the first embedding network. The flow is then conditioned on the output of the second embedding network.
Fig.~\ref{ml: fig: our method and embedding networks} shows a schematic overview of the model architecture.   

The first embedding network summarizes each single-event posterior.\footnote{As input data of the first embedding network, we use the standardized posterior samples (subtracting the mean and dividing by the standard deviation of the respective variable). This standardization of the input data is a common practice in \gls{ml} and allows for faster convergence of the model.} This network takes as input data the collection of $\npost$ posterior samples of the component masses and the luminosity distance (following Sec.~\ref{sec: method}); that is a three-dimensional posterior distribution for the $\nbatchevents$ events in the sub-population. 
We have found that 16 ``summary'' parameters for the single-event posterior are sufficient to recover the population posterior. 
The first embedding network is \textit{identical} for each event. 
If the flow analyzes $\nbatchevents$ events, we thus have $16\times\,\nbatchevents$ scalars describing the input data after applying the first embedding network. These data are then further summarized by an additional embedding network, whose output feeds into the flow. We choose this second summary to have 64 and 256 parameters for the two models we train in the result section below.
The embedding network parameters (summarizing the input data) also implicitly appear in the loss (cf.~Eq.~\ref{methods: machine learning: eq: theoretical loss for model}) and are therefore optimized jointly with the parameters that define the flow transformation as discussed below.
The two embedding networks significantly reduce the number of free parameters in the model, leading to less overfitting.

As anticipated, the normalizing flow is conditioned on the output of the second embedding network and computes an approximation of the true population posterior $p(\Lambda|\setdataK)$. 
In general, the normalizing flow performs a transformation that maps the physical variables (here the \metaps{} $\yvb\deffrom \Lambda$) to an unphysical variable $\zvb$ that follows the normal distribution.\footnote{We follow the standard notation as used in the review of normalizing flows of \cite{KobyzevPAMI2020}} 
One can rapidly generate samples in $\Lambda\sim q(\Lambda|\setdataKi)$ by drawing samples from the normal distribution ($\zvb\sim\mathcal{N}(\mu = 0, \sigma = 1)$) and applying the flow transformation to them, i.e.~$\Lambda = \gvb(\zvb)$.

We use the \texttt{nflows} \cite{nflows} package to construct this transformation, where the flow transformation from $\zvb$ to $\yvb$ is constructed from a sequence of simple transformations. 
In our case, these are piecewise rational quadratic coupling transforms \cite{NEURIPS20197ac71d43} in analogy to those implemented in \cite{Green:2020dnx}.\footnote{See app.~\ref{ml: results: app: coupling flows} for additional details on these transformations.}
The number of coupling transforms is referred to as the number of \textit{flow steps}. 
The parameters governing the coupling transforms are trained to minimize the loss defined in Eq.~\eqref{methods: machine learning: eq: theoretical loss for model}.
We have investigated different choices of parameters and found that four flow steps, each parameterized by a fully connected residual network with $32$ parameters and five to fifteen layers provided the best results. 
Table~\ref{tab: summary networks} summarizes the details of the specific network architecture. 
Throughout, we use \glspl{gpu} to accelerate both the generation of single-event posterior samples and the training of the normalizing flow.

\section{Astrophysical and instrumental setup}
\label{sec: astrophysical setup}

In the following we describe our assumptions on the astrophysical population of \glspl{bbh}, the detector network, the detection criterion and the generation of waveforms. 

\subsection{Assumptions on the population distribution}

Throughout this work, we model sources as uniformly distributed in comoving volume in a flat $\Lambda$CDM universe described by the Hubble constant, $H_0$, and the matter content, $\Omegam$. We fix the latter to $ \Omegam  = 0.3$ and assume it to be known. This assumption is straightforwardly relaxed, but this is beyond the scope of our work here. 

The two training datasets we construct follow the \plmassname{} source frame mass distribution.\footnote{This is the simplest of the four source frame mass distribution considered currently by the LIGO-Virgo-KAGRA collaboration \cite{LIGOScientific:2018jsj, LIGOScientific:2020kqk, KAGRA:2021duu}. }
This source frame mass model is characterized by four parameters. The minimum mass $\mmin$ and the maximum mass $\mmax$ limit both source frame masses from below and above, respectively. In addition, we have two power law slope parameters $\alpha$ and $\beta$ characterizing the distribution of masses according to 
\begin{align}
    \label{eq: definition truncated PL mass model}p(\msone|\Lambda_m) &=\mathcal{N}\msone^{-\alpha}\chi_{\left[\mmin,\mmax\right]}(\msone)\,,\\\nonumber
    p(\mstwo|\msone,\Lambda_m)
    &=
    \mathcal{N}'\mstwo^{\beta}\chi_{\left[\mmin,\msone\right]}(\mstwo)\,,
\end{align}
where we have defined the normalization constants $\mathcal{N}$ and $\mathcal{N}'$, as well as the set of \metaps{} $\Lambda_m = \{\mmin, \mmax, \alpha, \beta\}$. Finally, $\chi$ is the characteristic function, defined as 
\begin{equation}
    \chi_{\left[a,b \right]}(x) 
    \deffrom
    \begin{cases}
        1 & \text{if } x \in \left[ a,b\right] 
        \\
        0 & \text{otherwise.}
    \end{cases}
\end{equation}

\subsection{Assumptions on BBH sources}

We focus on precessing \glspl{bbh} in quasi-circular orbits, characterized by 15 parameters: the component (detector frame) masses of the BHs, $\mdone, \mdtwo$, the two spin vectors described by their magnitudes, $a_1, a_2$, angles of reference, $\phi_{12}, \phi_{jl}$, and tilts, $\theta_1,\theta_2$, as well as the time of coalescence, $t_c$, the phase at coalescence, $\Phi_c$, declination and right ascension of the sky position, $\delta_{\rm SP}$ and $\alpha_{\rm SP}$, luminosity distance, $d_L$, the inclination of the orbital plane with respect to the line of sight, $\theta_{jn}$ and polarization angle, $\psi$. 
We assume sources to be distributed uniformly in the sky, we draw the spins isotropically over the sphere and the spin magnitudes uniformly between 0 and 0.99.

The \phXPHM{} waveform \cite{Pratten:2020ceb} is used to model the gravitational wave signal of the \gls{bbh} coalescences in the frequency domain.

\subsection{Assumptions on waveform and the detector network}


We assume the O1 sensitivity curve \cite{det1-aligo2015} for the \gls{ligo} Hanford and \gls{ligo} Livingston detectors and impose as a selection criterion a \gls{snr} threshold of 12.\footnote{\newcom{This is an approximation, since the application of this method to real data will make the selection criteria more intricate, e.g.~incorporating the false alarm rate. To fully account for selection effects, an injection campaign would be needed. }}
Given a detected \gls{gw} signal, we then use the deep learning tool \dingo{} \cite{Dax:2021myb, Dax:2021tsq} to analyze the strain data and produce single-event posterior samples. 
Due to limitations in the size of the parameter space of signals that can be reliably analyzed we restrict the mass range of signals to a conservative mass cut-off \textit{in source frame mass} of $\msone\geq\mstwo\geq 18\,\msun $.\footnote{In the near future, the lower bound of \dingo{}'s mass range is expected to decrease to 5\,$\msun$. } This lower bound on the mass range directly implies a lower bound on the prior of $\mmin$ we can explore. 
Also, recall that the mass spectrum method uses only the component masses and luminosity distance. We thus discard the remaining single-event parameters. 
See table~\ref{tab: summary networks prior} for the priors on the training set and, by extension, the prior learned by the model.

\section{Training datasets}
\label{sec: training sets}

The number of events per sub-population, $\nbatchevents$, is a free parameter of this approach, the optimal value of which we would like to determine. 
To study this, we build two different training sets and train models on each one, assuming the same detector network and with the population model, described in Sec.~\ref{sec: astrophysical setup}.
In the following, we refer to a \textit{\datasetsample{}}, as one population of events that share the same \metaps{}. 
Training set \lowevents{} has $\trainingsetonenumberPL{}$ \datasetsamples{}, which is ten times more \datasetsamples{} than that of training set \highevents{} (which has $\trainingsettwonumberPL{}$ \datasetsamples{}). However, training set \lowevents{} has only ten events per population, which is 20 times less than training set \highevents{} (which has $200$ events per \datasetsample{}). 
Overall, the two datasets contain approximately the same number of \gls{gw} events (and posterior samples) and hence, their information content (and their computational cost) is also approximately equal.
For this reason, the performance of the models trained on the respective datasets can be directly compared. 
Training dataset \lowevents{} only allows small event sub-populations ($\leq 10$), but with an in-depth training on many population examples, whereas training dataset \highevents{} allows for large sub-populations at the price of a limited number of populations.

\subsection{Training set \lowevents{}}

Each \datasetsample{} of the training data contains the true value of the \metaps{}, $\Lambda$, ten events and 200 associated posterior samples in three variables: the (detector frame) component masses and luminosity distance. In total, the data associated to one population \datasetsample{} thus contains $6000=10\times200\times3$ scalars. 

During one training epoch, we randomly choose $\nbatchevents$ events among the ten events for each \datasetsample{} with $\npost$ random posterior samples each. This sampling method increases the variability of the input data. 
After several trials, we have found that the model provides a good approximation to the population posterior if it analyzes six events per sub-population ($\nbatchevents = 6$).\footnote{The combination of sub-populations of events delivers more reliable posterior distributions if the number of events per sub-population ($\nbatchevents$) is high. However, the time to generate the training dataset limits the maximum value $\nbatchevents$ which therefore, cannot be set arbitrarily high. } The same reasoning applies to the number of posterior samples, 100 posterior samples per event seem to be sufficient to produce a faithful approximation (although see the discussion below in Section~\ref{subsec: results training set 2}). 

A summary of the training dataset \lowevents{} can be found in table \ref{tab: summary training datasets}.

\subsection{Training set \highevents{}}

With future \gls{gw} detector networks in mind, we also construct a training dataset that allows for models that analyze a much larger number of events. Each population \datasetsample{} of the training set \highevents{} includes 200 events with 200 posterior samples per event. The data associated to each \datasetsample{} thus includes $120,000 = 200\times 200 \times 3$ scalars. The model we train below selects randomly 100 out of the 200 available events during each training epoch. For each of these events, the flow chooses 100 out of 200 posterior samples at random. Thus, each input population sample includes $30,000 = 100 \times 100 \times 3$ scalars. Again, this method of drawing a subset of the available data is introduced to reduce overfitting. 
For training set \highevents{} we find that $\nbatchevents = 100$ gives the best performing models. This is the result of a trade-off; given that one \datasetsample{} contains 200 \gls{gw} events, if $\nbatchevents$ is set higher the variability of the training data is not high enough, and if $\nbatchevents$ is too low we cannot analyze a large number of events, since too many sub-populations have to be combined. Empirically, we find that the model does not produce reliable results above $\mathcal{O}(10-20)$ combinations of different sub-populations.

\renewcommand{\arraystretch}{1.25}
\begin{table}[]
\centering
\begin{tabular}{ ccc } 
 \hline
 \hline
 \multicolumn{3}{c}{Summary of training datasets} \\
 \hline
 Study & \lowevents{} & \highevents{} \\
 \hline
 $\#$ training population samples &$\trainingsetonenumberPL{}$ &$\trainingsettwonumberPL{}$ \\
    $\#$ available events per population & 10 &200 \\ 
    $\#$ available posterior samples per event & 200 &200 \\ 
  \hline 
  \hline 
\end{tabular}
\caption{Properties of the two generated training datasets. }
\label{tab: summary training datasets}
\end{table}
\renewcommand{\arraystretch}{1.25}
\begin{table}[t]
\centering
\begin{tabular}{ ccc } 
 \hline
 \hline
\multicolumn{3}{c}{Summary priors} \\
\hline
Metaparameter & Prior & Unit \\
 \hline
 $H_0$ & $\mathcal{U}(40,140)$ & $\hu$\\
$\mmin$ & $\mathcal{U}(18,30)$ & $\msun$\\
$\mmax$ & $\mathcal{U}(37,47)$ & $\msun$\\
$\alpha$  & $\mathcal{U}(-2,2)$ & - \\
$\beta$  & $\mathcal{U}(-2,2)$ & - \\
 \hline
 \hline
\end{tabular}
\caption{Summary of priors assumed for the two training datasets. The uniform prior is denoted as $\mathcal{U}$. }
\label{tab: summary networks prior}
\end{table}

\subsection{Training the networks}

\begin{figure*}
  \subfloat[][]{\includegraphics[width = .47\linewidth]{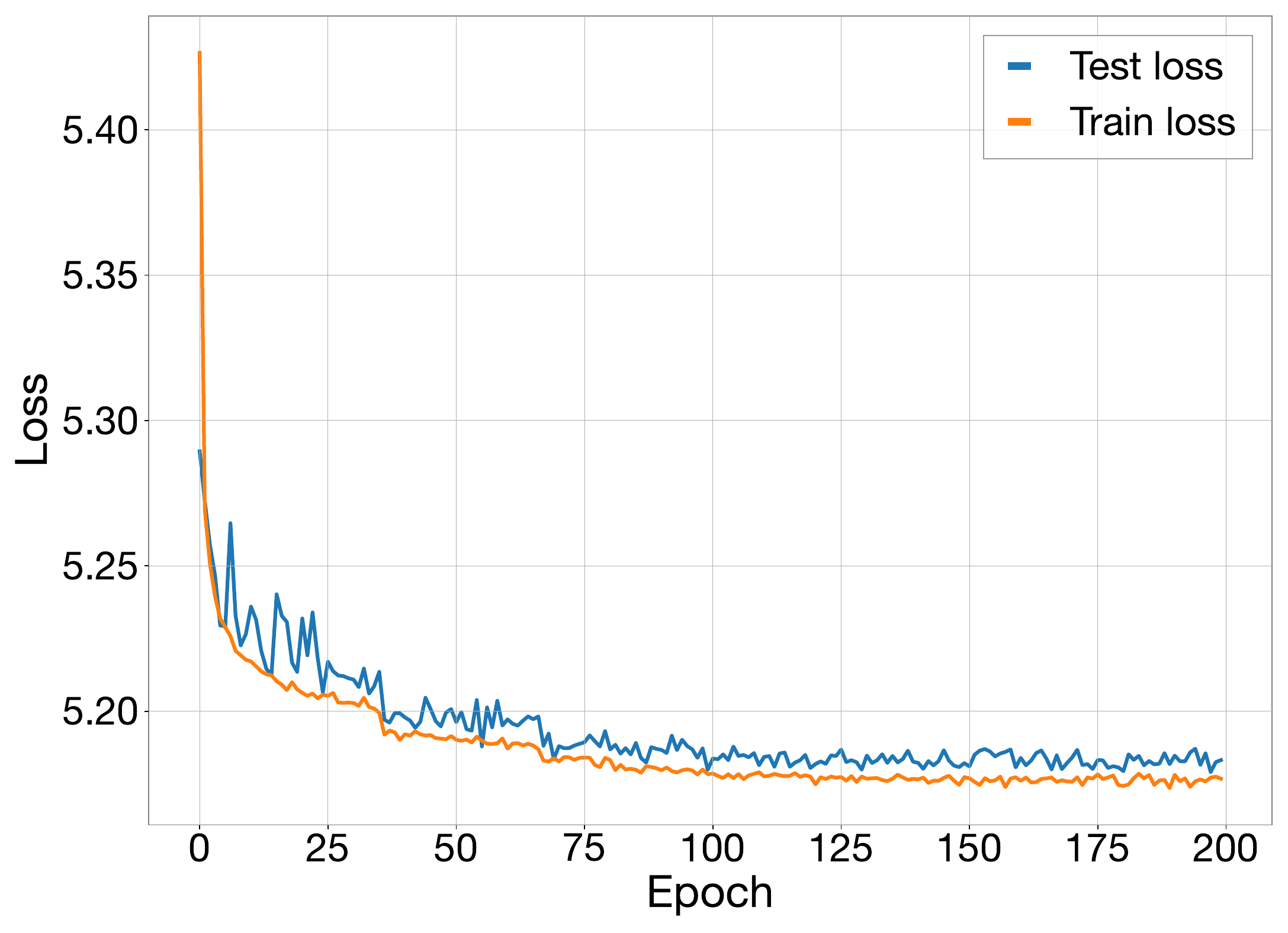}
  \label{fig: loss curves model 1}}
  \hfill
  \subfloat[][]{\includegraphics[width = .47\linewidth]{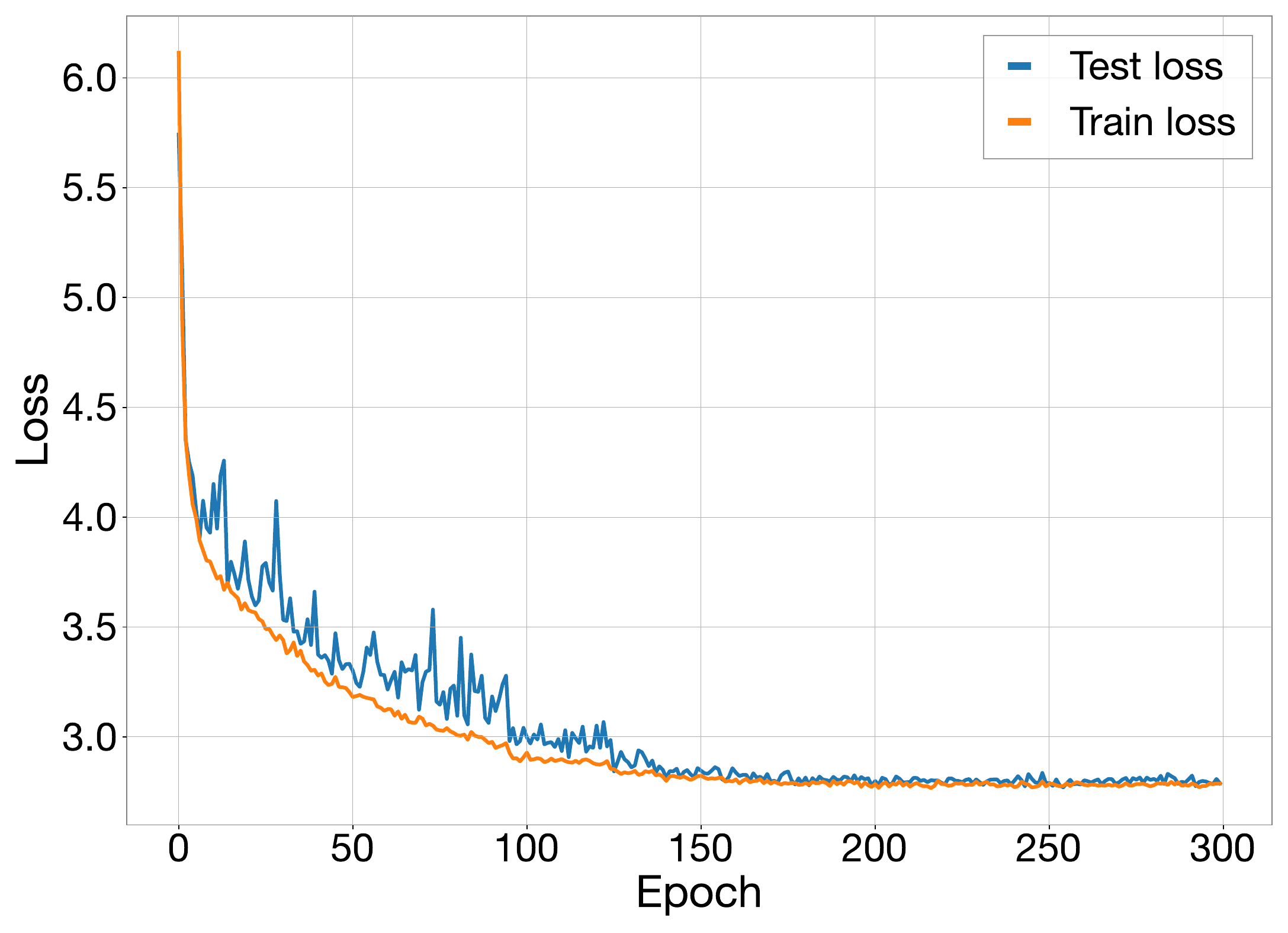}
  \label{fig: loss curves model 2}}
  \caption{Loss for the \textit{(left)} model \lowevents{} and \textit{(right)} model \highevents{}. Since the test loss (blue) and train loss (orange) do not differ much, we conclude that the models generalize well to unseen data.}
    \label{fig: loss curves}
\end{figure*}

Given the training data, we train different models by minimizing the loss we have introduced in Eq.~\eqref{methods: machine learning: eq: theoretical loss for model}, varying the network parameters, as well as the number of \gls{gw} events that are taken as input parameters. 
The parameters describing the flows that yield the best agreement with standard HBA results are summarized in table \ref{tab: summary networks}.
The network trained on dataset \lowevents{} (which we refer to in the following as model \lowevents{}) had a training time of $\sim5.5$~hours. 
In Fig.~\ref{fig: loss curves model 1}, we present the training and test loss curves for the model. Based on the loss curves, we conclude that the model can generalize effectively to data that were not included in the optimization process.

The training time of model \highevents{} was 82 minutes. The shorter training time (compared to model \lowevents{}) is due to the smaller number of \datasetsamples{} in training dataset \highevents{}. The associated training and test loss curves of model \highevents{} are plotted in Figure~\ref{fig: loss curves model 2}. 
We discuss the resulting respective posterior distributions in Sec.~\ref{subsec: results training set 1} for the training set \lowevents{} and Sec.~\ref{subsec: results training set 2} for training set \highevents{}. 

\section{Results}
\label{sec: results}

\subsection{Results with model \lowevents{}}
\label{subsec: results training set 1}

\renewcommand{\arraystretch}{1.3}
\begin{table}[t]
\centering
\begin{tabular}{ ccc } 
 \hline
 \hline
 \multicolumn{3}{c}{Summary of the normalizing flow parameters } \\
 \hline
 \diagbox[]{Variable}{Model} & \lowevents{} & \highevents{} \\
 \hline
\specialcell{Events per\\batch $\nbatchevents$}  & 6 & 100  \\ 
\specialcell{Posterior samples\\per event $\npost$} &  100 & 100 \\ 
\hline 
\specialcell{Dimensions embedding\\network 1} & \specialcell{ (512, $256^5$,\\ 128, 64)} & \specialcell{(512, 256, \\128, 64)} \\ 
\specialcell{Dimensions embedding\\network 2} &  \specialcell{($512^4$, $256^4$,\\ $128^2$, $64^3$)}  & \specialcell{($1024^2$, $512^2$,\\ 256)} \\ 
Flow steps & 3 & 4 \\ 
Spline points    & 8 & 6 \\ 
\specialcell{Hidden dimensions\\(spline network)}& 32 & 32 \\ 
\specialcell{Hidden layers\\(spline network)} & 5 & 15 \\ 
\hline 
Training epochs & 200 & 300 \\ 
Learning rate  &  0.0001 & 0.0001 \\ 
Scheduler &Plateau&Plateau\\ 
Batch size &  1024 & 1024 \\
 \hline
 \hline 
\end{tabular}
\caption{Architecture of the embedding networks and the normalizing flow. For the hidden layers of the embedding networks we use the tuple notation $X^n \deffrom (X,X,\ldots,X)$, with $X$ repeated $n$ times. }
\label{tab: summary networks}
\end{table}

\begin{figure*}[]
  \subfloat[][]{
    \includegraphics[width = .48\linewidth]{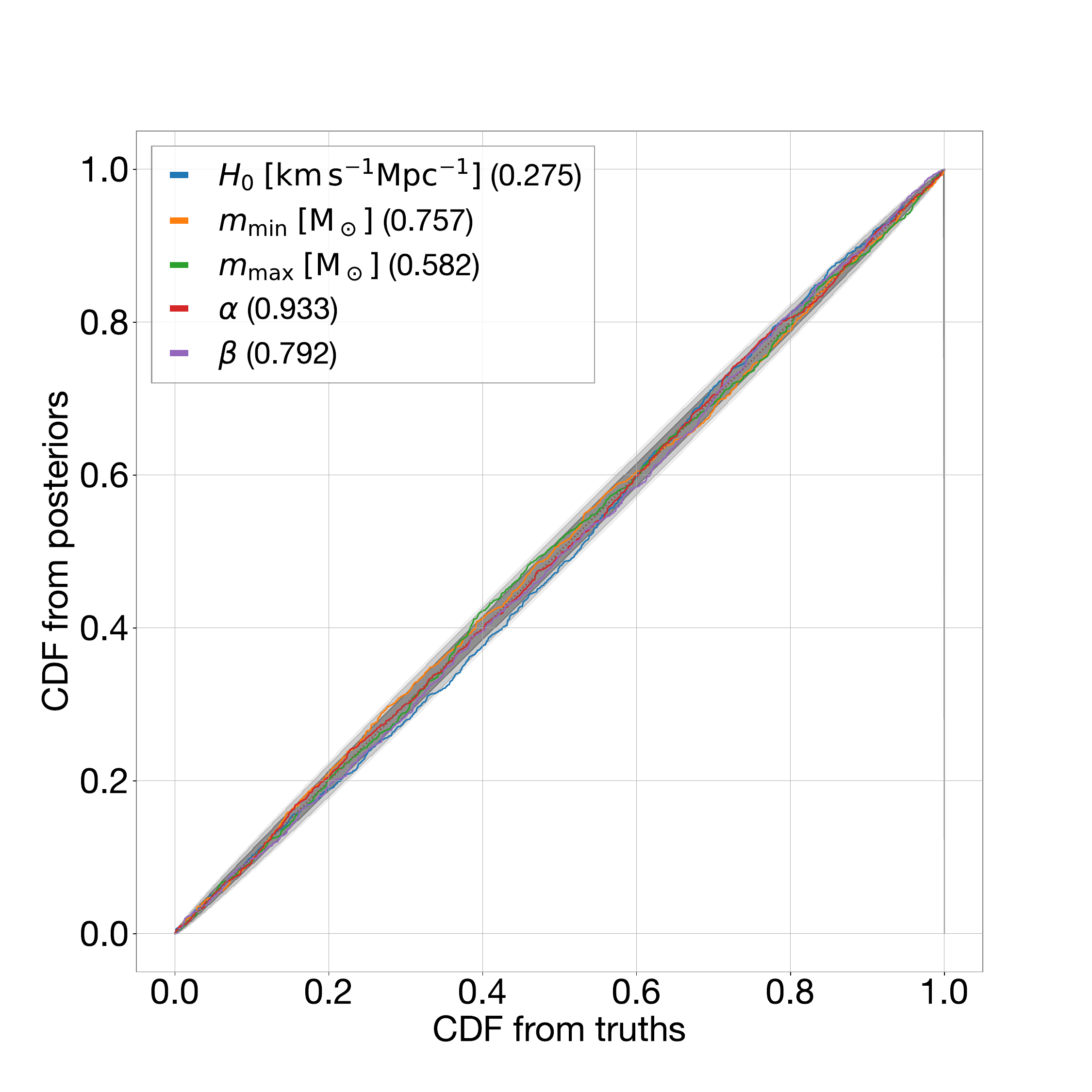}
    \label{fig: p-p plot model 1}
    }
  \hfill
  \subfloat[][]{
    \includegraphics[width=0.48\textwidth]{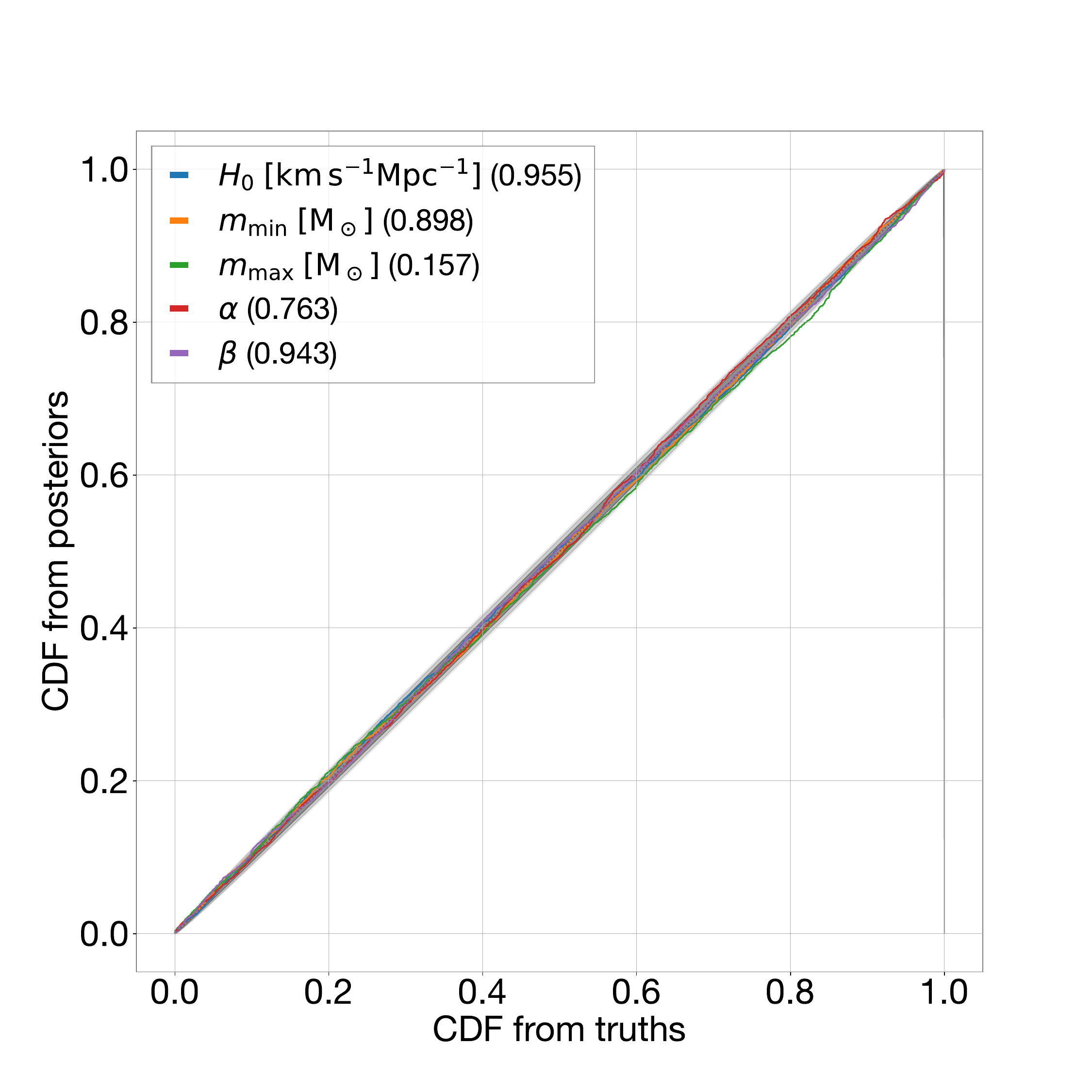}
    \label{fig: p-p plot model 2}
    }
    \caption{The P-P-plot for 1000 injections for model \lowevents{} (\textit{left}) and 2500 injections for model \highevents{} (\textit{right}). We find p-values as indicated in the legend, indicating that the models reconstruct the population posterior correctly. However, the lowest p-values of model \highevents{} is slightly lower than the model \lowevents{}. The $\mmax$ parameter of model \highevents{} has the lowest value, with $15.7\%$.}
    \label{fig: pp plots}
\end{figure*}

As a first validation step, we generate the P-P-plot of the model. 
To this end, we draw 1000 population \datasetsamples{} from the training dataset, input the corresponding posterior samples in the model, and sample $q(\Lambda|\setdataKi)$ for each. From the $\Lambda$ samples, we compute the percentile in which the true value of the population lies and sort the resulting percentiles by value. The cumulative density of the percentiles is shown in Fig.~\ref{fig: p-p plot model 1}. If the model correctly infers the \metaps{}, the figure should follow the diagonal within a reasonable error interval. 
From the \gls{ks} test (comparing the computed percentiles against the uniform distribution), we obtain p-values between $27\%$ and $93\%$, which is expected for five variables, indicating that the model reconstructs the population posterior correctly. 

The above tests only include $\nbatchevents = 6$ events per population. 
To further validate our results, we generate a detected population with $\resultsonenevents$ \gls{gw} events. With \dingo{}, we produce posterior samples for each of these events and run a classical \gls{mcmc} analysis (with an analytical likelihood, using \icarogw{} \cite{Mastrogiovanni:2021wsd}) on these samples. The resulting \metap{} posterior distribution serves as our ground truth. 
This scheme differs from the classical approach (for instance in \cite{KAGRA:2021duu, LIGOScientific:2021aug}) as the computationally expensive \gls{pe} has been carried out with \dingo{}. 
This ``crossing'' from the likelihood-free inference part to the classical inference (with \icarogw{}) is a simplification and comes at a cost. 
If \dingo{} does not correctly estimate the single-event posterior distribution, the resulting distribution of the \metaps{} with \icarogw{} will not represent the ground truth. We will discuss below for model \highevents{} the possible consequences of this assumption.
Note that it is possible to correct for possible inaccuracies of \dingo{} by importance sampling. However, since this significantly increases the computation time we do not choose to pursue this here. 

\begin{figure*}
    \centering
    \includegraphics[width=0.85\textwidth]{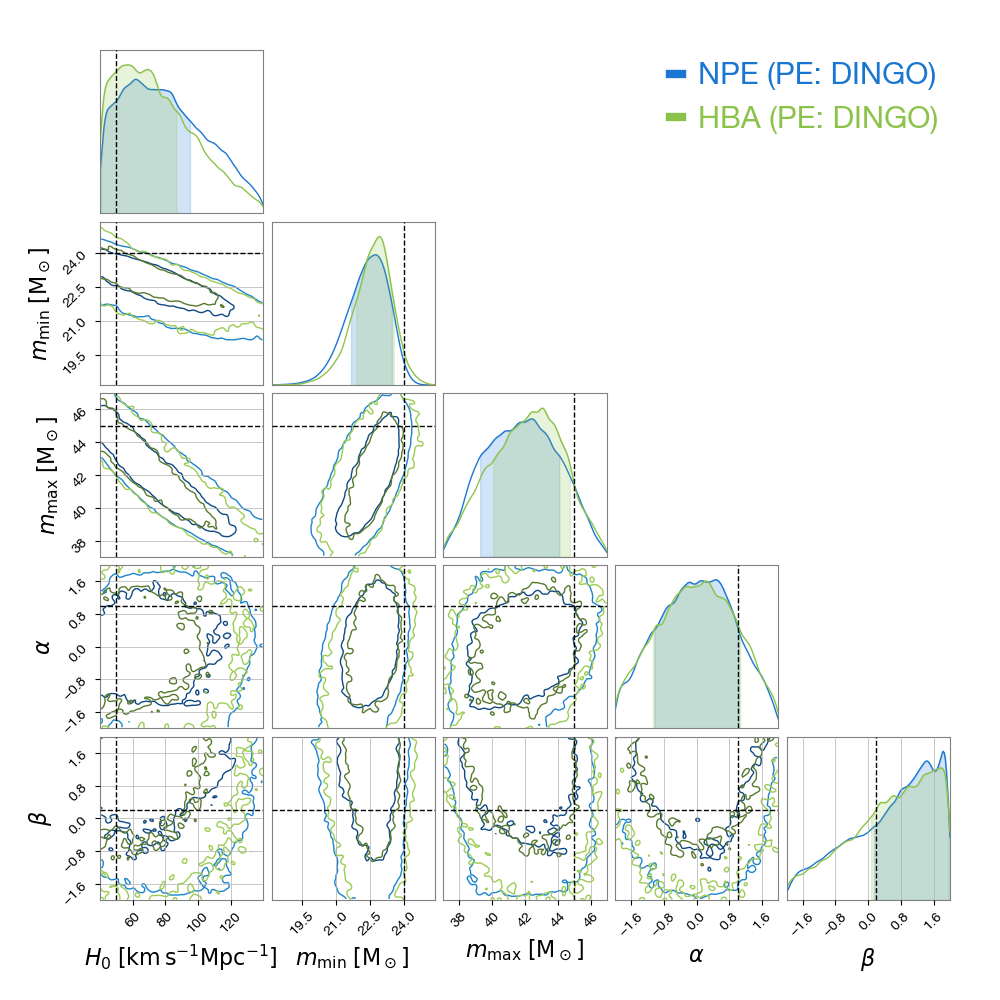}
    \caption{Results from model \lowevents{} (NPE, blue) compared to a conventional \acrlong{hba} (HBA, green). The posterior analyzes $\resultsonenevents$ \gls{gw} events of population no.~4 (cf.~app.~\ref{app: summary populations}). The one and two sigma intervals are indicated as two-dimensional contours and dashed lines mark the true values of $\Lambda$. We could produce an effective sample size of $\sim8\times 10^4$ posterior samples in $2.5$~minutes of computation time. In total, we have verified our results on a total of twelve populations. For the largest discrepancy between our model and the conventional approach consider the result in Fig.~\ref{fig: importance sampling}. The ``PE'' stands for individual event parameter estimation which, in both cases, uses the \dingo{} algorithm. }
    \label{fig: combination events model 1}
\end{figure*}

In parallel, we apply our model to the same single-event posterior samples and combine the results with the importance sampling step that was outlined in Sec.~\ref{methods: machine learning: subsec: combining events}. This procedure is applied for twelve different populations.\footnote{Each population has a different Hubble constant, and variables parametrizing the source frame mass distribution. The details of the populations are given in tab.~\ref{tab: summary populations}. } 
Fig.~\ref{fig: combination events model 1} shows one out of these twelve distributions, with a model that analyzes $\resultsonenevents$ events in total. Since the network was trained for $\nbatchevents = 6$ events, we divide the input data in $\nbatch = 10$ sub-populations.
The result shows that it is possible to combine the output of multiple model evaluations and obtain the correct population posterior.
This figure is representative of the majority of cases -- we generally see good agreement between the two methods. 
We have also verified that the (arbitrary) division of events in the different sub-populations does not impact the resulting population posterior. 

To make this comparison more quantitative, we compute the \gls{js} divergence\footnote{The \gls{js} divergence is a symmetrized version of the \gls{kl} divergence that was defined in Eq.~\eqref{methods: machine learning: eq: def: kullback leibler divergence}. } between the \gls{npe} and \gls{hba} results for each of the variables in $\Lambda$. Tab.~\ref{tab: js divergence} collects these values. The lower the \gls{js} divergence, the better the agreement between two distributions. The \gls{js} divergence for single-event \gls{pe} with \lalinference{} (for identical runs with different random seeds) is $\sim 7\times 10^{-4}\,\natunit$ \cite{Romero-Shaw:2020owr}. For two \icarogw{} runs with the same settings, we find \gls{js} divergences between $3\times 10^{-3}\,\natunit$ and $10^{-4}\,\natunit$.  
The \gls{js} divergences observed in our experiments are an order of magnitude higher than these baselines. Thus, there exists potential for further refinement in our approach. 
The problems appear almost exclusively for two \metaps{}: the minimum and maximum mass of the population, $\mmin$ and $\mmax$, and out of the two, the minimum mass proves to be the more difficult parameter. Out of the 60 \gls{js} divergences we have analyzed (twelve populations with five \metaps{} each), 16 had a \gls{js} divergence larger than 0.01. The population models which proved to be most difficult to reconstruct were populations~1, 7 and 10, where population 1 and 7 both have low Hubble constant. Additionally, population 7 has $\mmax = 38\,\msun$, very close to the boundary of the prior for which the model was trained ($\mmax = 37\,\msun$). It is well known the neural network performance decreases close to the prior boundary.  
We show in the next section that it is possible to recover the \gls{hba} result by applying importance sampling to the samples produced from our model. 

\begin{table}[]
    \centering
    \begin{ruledtabular}
        \begin{tabular}{cccccc}
        \multirow{2}{*}{Population} & \multicolumn{5}{c}{JS divergence ($10^{-3}$~nat)} \\
         &  $H_0$ &  $m_{\rm min}$ &  $m_{\rm max}$ &  $\alpha$ &  $\beta$ \\
        \hline
            0  &  6.4 &  26.4 &   4.9 &    2.9 &   1.6 \\
            1 (Fig.~\ref{fig: importance sampling}) & 10.0 &  70.9 &  15.3 &   17.7 &   4.0 \\
            2  &  3.8 &   6.3 &  11.9 &    4.1 &   5.8 \\
            3  &  3.2 &   4.3 &   4.2 &    8.5 &   1.5 \\
            4 (Fig.~\ref{fig: combination events model 1})  &  7.0 &  10.3 &   3.5 &    0.9 &   1.4 \\
            5  &  0.9 &   4.2 &   7.2 &    6.8 &   0.8 \\
            6  &  1.8 &   6.6 &  11.6 &    8.1 &   3.5 \\
            7  &  2.6 &  16.4 &  18.4 &    3.4 &  21.3 \\
            8  &  8.9 &   4.6 &   4.5 &    3.4 &   1.0 \\
            9  &  2.3 &  10.1 &  17.2 &    6.2 &   2.1 \\
            10 & 30.6 &  10.2 &  21.8 &    4.0 &   0.8 \\
            11 &  3.4 &   4.1 &  11.1 &    6.9 &   2.6 \\
        \hline
            Mean &  6.73 & 14.53 & 10.97 & 6.07 & 3.86 \\
            Median &  3.61 & 8.36 & 11.35 & 5.16 & 1.81 \\
        \end{tabular}
        \end{ruledtabular}
    \caption{\gls{js} divergence (in units of $10^{-3}\,\natunit$) for all \metaps{} and all twelve populations.
    The mean and median values over all populations are also presented. 
    The most problematic parameters are $\mmin$ and $\mmax$. We have indicated which population are plotted in later sections, corresponding to cases where our model differs weakly or strongly to the conventional approach. }
    \label{tab: js divergence}
\end{table}

\subsubsection{Failing of the model and recovery from importance sampling}

\begin{figure*}
    \centering
    \includegraphics[width=0.8\textwidth]{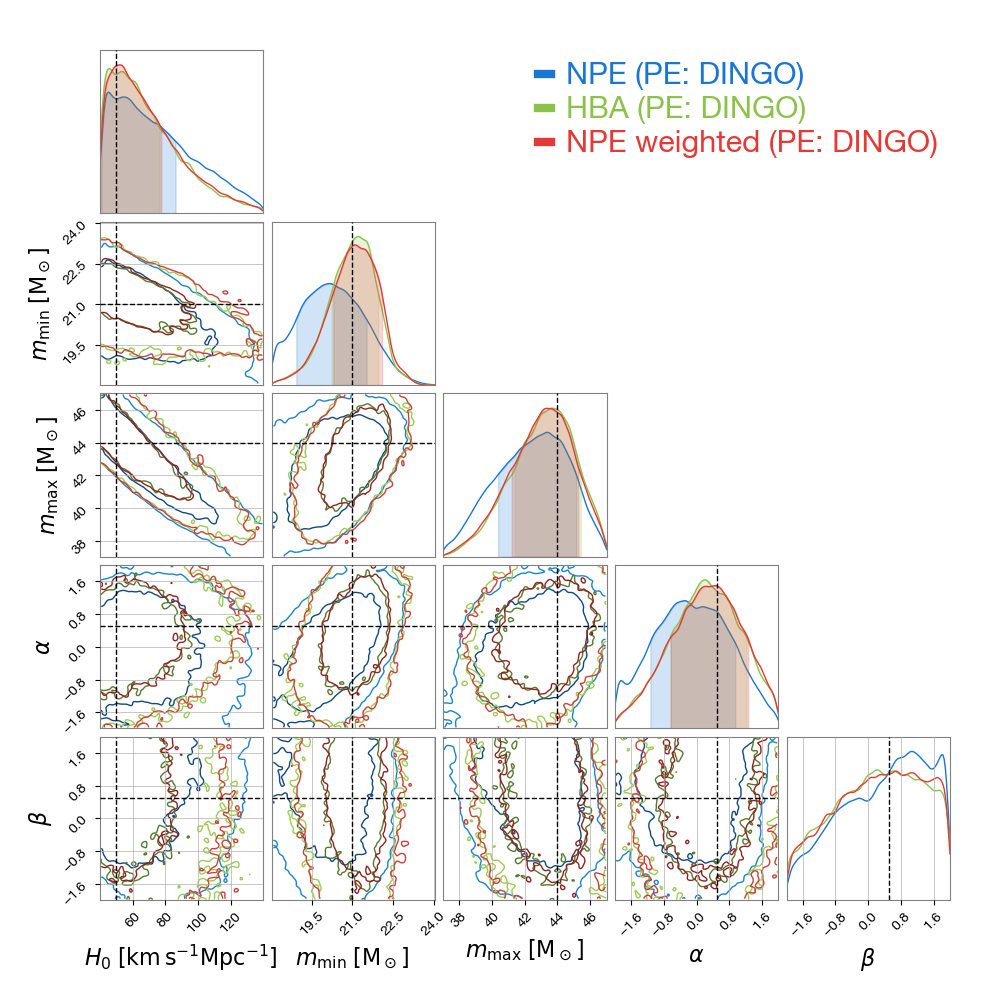}
    \caption{Results from model \lowevents{} (NPE (PE: DINGO), blue) compared to a conventional \acrfull{hba} (green). The posterior analyzes $60$ \gls{gw} events of population no.~1 (cf.~app.~\ref{app: summary populations}). The one and two sigma intervals are indicated as two-dimensional contours and dashed lines mark the true values of $\Lambda$. As the \gls{nf} and the classical analysis differ, most notably in the variable $\mmin$, we perform an importance sampling with the classical likelihood. The resulting posterior is shown in red (NPE (weighted)) and agrees well with the classical result.}
    \label{fig: importance sampling}
\end{figure*}

In certain cases the \gls{npe} samples do not agree with the \gls{hba} samples. 
However, we have access to the probability associated with which each population sample was generated through the construction of the \gls{nf}. 
We can therefore obtain the \gls{hba} result by calculating the (conventional) population likelihood for each sample, $p(\Lambda|\setdataC)$, and reweighting the \gls{nf} samples to this target likelihood, using the weights determined by Eq.~\eqref{eq: is weights classical recovery}. 
Fig.~\ref{fig: importance sampling} shows this procedure on the example of population 1, and model \lowevents{}. 
\newcom{For comparison, the current LVK cosmological inference code produces the classical result (with $24,000$ samples and parallelizing on 16 cores) in $\sim8$~hours}, whereas the flow produced 300,000 samples in $2.3$~minutes. Applying an additional importance sampling step (parallelizing on 16 cores) generated an effective sample size of $10^4$ in $\sim 3.3$~hours. This gain in computation time is due to the reduced number of likelihood evaluations with the \gls{npe} ($3\times10^5$) when compared to the \gls{hba} ($1.7\times10^6$).
These computation times do not include the times for single-event parameter estimation (here carried out with \dingo{} and therefore, within minutes).
\newcom{Also note that one can implement hierarchical inference codes with just-in-time compilation, using \glspl{gpu} and faster sampling algorithms \cite{Edelman:2022ydv} that generate \metap{} posteriors in minutes. }

When combining more than $ 10-20$ sub-populations of events, the resulting \gls{npe} posterior becomes unreliable.\footnote{\newcom{We occasionally find a non-smooth posterior distribution for models that analyze one sub-population. While this does not significantly impact the result, these discontinuities accumulate when we compute the product of several posteriors that each analyze one sub-population, respectively.
This is likely a consequence of the chosen model architecture and not intrinsic to the method. }}
Consequentially, model \lowevents{} cannot analyze more than $\sim 100$ events.
This might be caused by model \lowevents{} not resolving the fine structure in the posterior distribution that becomes important when combining large event sets. 
We thus rely on training set \highevents{} to construct a model that can analyze a larger number of events as we now elaborate.

\subsection{Results with model \highevents{}}
\label{subsec: results training set 2}

To show the capability of the model to reconstruct the population posterior given a large number of observed events, we use training set \highevents{} to construct a model analyzing $\nbatchevents = 100$ \gls{gw} events. 
Fig.~\ref{fig: loss curves model 2} shows the resulting loss curves of the training and test dataset, respectively. 
The train and test loss coincide, suggesting that the model can process unseen input data and generate accurate hyperparameter posterior distributions.
Fig.~\ref{fig: p-p plot model 2} shows the P-P-plot of model \highevents{} with 2500 population realizations, implying that the network has correctly learned the desired posterior distribution. 

\begin{figure*}
    \centering
    \includegraphics[width=0.95\textwidth]{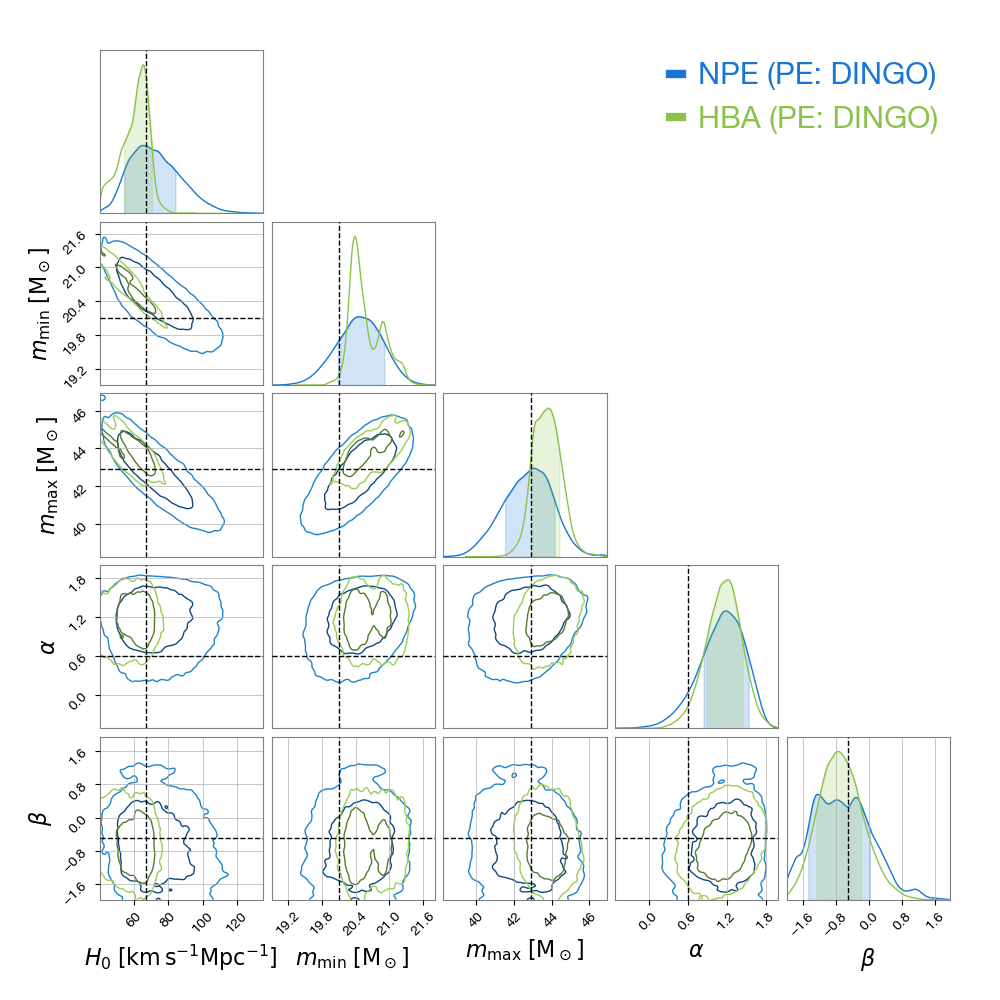}
    \caption{Results from model \highevents{} (blue) compared to a conventional \acrfull{hba} (green). 
    Both analyses use the \dingo{} samples as input data. 
    The posterior analyzes $\resultstwonevents$ \gls{gw} events. The one and two sigma intervals are indicated as two-dimensional contours and dashed lines mark the true values of $\Lambda$.
    The model and the conventional analysis show a discrepancy. However, the model is generally closer to the injected values than the \gls{hba} method that relies on \dingo{} \gls{pe} samples. 
    }
    \label{fig: combination events model 2}
\end{figure*}

We have verified that when analyzing 100 \gls{gw} events that the \gls{nn} is in good agreement with the \gls{hba} for all the populations described in table~\ref{tab: summary populations}. 
Considering a larger number of events, we focus on one specific population, with the parameters $H_0= 67\,\hu, m_{\rm min}= 20.1\,\msun, m_{\rm max}= 42.9\,\msun,  \alpha= 0.6$ and $\beta= -0.5$. 
Fig.~\ref{fig: combination events model 2} compares the posterior of our model and the classical posterior, analyzing 600 events. Although the posterior distributions overlap, they show a significant deviation.\footnote{As previously, we can  successfully recover the \gls{hba} result through importance sampling.} The computation time for the conventional approach was 127~hours\footnote{This computation time was for an injection set (used to compute the selection effect) of $1.4\times 10^5$ detected \gls{gw} signals. 
} and for our model 7~minutes.

The reason for the discrepancy in Fig.~\ref{fig: combination events model 2} is still an open question: 
as anticipated above, we make an approximation of the ground truth. We use \dingo{} to estimate the single-event posterior distributions, and these samples are then subsequently analyzed by \icarogw{} to derive the \metap{} posterior. To decrease the computation time we did not perform importance sampling on the generated \dingo{} samples for the individual event posteriors. This can degrade the performance of the estimation of the single-event posterior. 
From a preliminary analysis, we find that some posterior distributions as inferred by \dingo{} differ from the \gls{pe} samples of \bilby{}. 
In the near future, we hope to perform a full \gls{pe} on all events and compute the ground truth from conventional analyses alone.
However, we emphasize that even if the \dingo{} algorithm is not a perfect approximation to the single-event posterior, this does not invalidate our approach -- by construction the \gls{npe} model learns the posterior distribution marginalized over the \dingo{} uncertainty. 

There are other potential sources of the discrepancy, such as the smaller number of \datasetsamples{} represented in the training dataset \highevents{}. We have checked that the posterior resulting from a model trained with training set \lowevents{} is compatible with a posterior trained with training set \highevents{}.  

Moreover, we find a strong dependence of the \gls{hba} results on the number of posterior samples per event (if the number of posterior is not ``high enough''). This potentially additional source of uncertainty is now discussed.


\subsubsection{Impact of the number of posterior samples}

The \gls{hba} scheme usually processes $\mathcal{O}(10^3-10^4)$ posterior samples per event. Since the \gls{npe} model works with a significantly lower number, this section explores the consequences of this approximation.

\begin{figure*}
    \centering
    \includegraphics[width=0.75\textwidth]{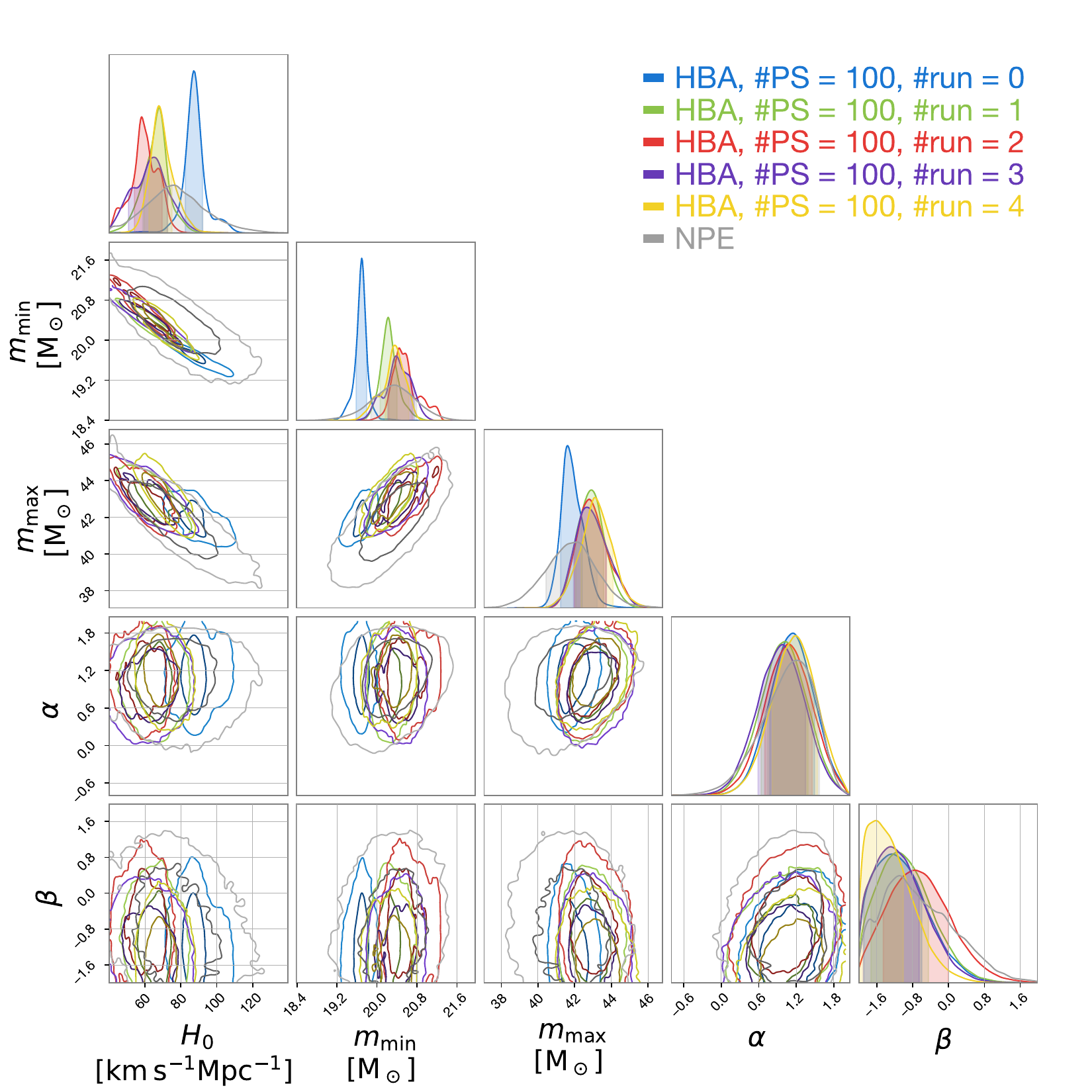}
    \caption{Results from repeating the \gls{hba} analysis with different random sets of 100 samples from each single-event posterior. We compare this to the \gls{npe} scheme, plotted in grey. 
    The posterior here was computed from 300 events, and the true population parameters are indicated by the dashed lines. 
    }
    \label{fig: impact number of posterior samples}
\end{figure*}

We carry out the \gls{hba} using 100 posterior samples per event, for 300 events in total. The population analyzed is the same as in the previous paragraph.
We repeat the \gls{hba} many times, using a different set of posterior samples for each event each time. This leads to a scatter of the population posterior, as shown in Fig.~\ref{fig: impact number of posterior samples}.

For the hyperparameters with more scatter (in particular $H_0$ and $\mmin$) the \gls{npe} differs more from the individual \glspl{hba} and gives posteriors that are broader, covering the range over which the individual \glspl{hba} vary. So, 
the \gls{npe} is mass-covering, i.e.~it has support across the scatter of the posterior that arises from the limited number of posterior samples. 
This indicates that the \gls{npe} marginalizes over the additional uncertainty arising from this approximation. This behavior is also expected from the construction of the loss in Eq.~\eqref{methods: machine learning: eq: loss rewritten 2 expectation values} -- if the model $q(\Lambda|\setdataK)$ has no support on the support of $p(\Lambda|\setdataK)$, the loss diverges.
As a consequence, future analyses will either have to increase the number of posterior samples or use a more complete summary of the \gls{gw} signal. In this limit, we expect to find a closer agreement between these two methods.\footnote{\newcom{Note that the works of \cite{Golomb:2022bon, Talbot:2023pex} have explored these consequences for conventional analyses. An insufficient number of posterior samples per event was found to lead to narrow, incorrect population posteriors. }}

\section{Conclusion}
\label{sec: conclusion}

Future planned detector networks will detect up to a hundred thousand GW sources each year, allowing for high-precision measurements of the parameters characterizing the population, including cosmological parameters. 
With these many events, fast methods, such as machine learning, will be essential for population inference and related analyses, e.g., tests of \gls{gr}.

In this work, we have demonstrated that normalizing flows can rapidly produce the posterior distribution of cosmological and population parameters inferred from observed \gls{gw} dark sirens. 
We have introduced a loss function one has to maximize for the \gls{nf} to converge to the true posterior distribution.
Within this setup, the posterior learned naturally incorporates selection effects. 
Normalizing flows prove to be flexible enough to approximate the posterior distribution of up to 600 \gls{gw} events. 

However, there are instances where the results from the flow do not align with the standard results from the \gls{hba}. 
\newcom{The work of \cite{Golomb:2022bon, Talbot:2023pex} has shown for conventional HBAs that incorrect population posteriors can arise due to insufficient samples per Monte Carlo integral, an effect which could contribute to the observed differences. }
An almost perfect agreement can still be obtained by performing importance sampling on the samples outputed by the neural network, using the standard HBA likelihood for the weights. 
This process increases the computational time of our method, but still requires $\mathcal{O}(10)$ fewer likelihood evaluations than in the standard HBA approach. 

The reasons for the discrepancies between HBA and our method are still unresolved.  
The single-event posterior samples generated by \dingo{} could deviate from the true single-event posteriors, possibly leading to biases in the HBA results.
Our method relies on an arbitrary data summary (provided the summary contains the majority of the signal information).\footnote{The choice we made here, to represent data by physical posterior samples, $\theta$, is not the only possibility. For instance, another approach would be to directly compress the strain data. This compressed data can then provide the input for the population inference with a \gls{nf}. As such, our approach is particularly adapted to a population analysis relying on other \gls{nn} summaries. }
As long as the flow trains on this summary data, the model should recover the true \metap{} posterior. Indeed, the network could, in principle, compensate for the eventual incorrect representation of the GW data, but we have not explicitly shown this in the present work. 
Moreover, we currently use 100 posterior samples per \gls{gw} event which also leads to an additional uncertainty. 
To produce reliable posterior distributions one has to use a sufficient number of posterior samples per event, posing a potential bottleneck for analyses of 3G detector data. Indeed, future analyses might have to limit the number of posterior samples for computing efficiency, resulting in an additional uncertainty our approach can marginalize over.
Alternatively, a possibility is that the model has not accurately learned the population posterior, but we have performed several tests that make this scenario unlikely.

Our approach requires to divide the observed population into sub-populations of fixed dimensionality to use as input for the network. Errors in the learning of the posterior accumulate when combining the results of sub-populations. In practice, we find that when the model is repeatedly evaluated to combine a large number ($\mathcal{O}(10-20)\times\nbatchevents$) of sub-populations of events, instabilities appear that prevent the production of an accurate posterior, i.e.~\gls{gw} catalogs with $\nobs / \nbatchevents \gtrapprox 10$ cannot be robustly analyzed with the current framework. A more robust network architecture might be needed to learn the posterior with the necessary accuracy to analyse $\mathcal{O}(1000)$ events. We leave this for future work, as well as more complex mass and redshift distributions.

Finally, the method proposed can test population models with source frame mass distributions that are difficult to parametrize analytically since it relies on simulation-based inference, with (in principle) no explicit likelihood needed. For instance, one could include stellar evolution codes, circumventing the choice of a analytic distribution describing the source frame mass distribution.

\section*{Acknowledgments}

KL is grateful to the Fondation CFM pour la Recherche in France for support during his doctorate.
We thank Eric Chassande-Mottin for comments on the manuscript. 
Numerical computations were performed on the DANTE platform, APC, France. 
Numerical computations were partly performed on the S-CAPAD/DANTE platform, IPGP, France.
This material is based upon work supported by NSF's LIGO Laboratory which is a major facility fully funded by the National Science Foundation.

\appendix

\section{Coupling flows}
\label{ml: results: app: coupling flows}

We provide details of the deep \gls{nn} in this appendix. The flow transformation of Sec.~\ref{subsec: flow architecture}, $\gvb$, is given by a sequence of coupling transforms \cite{dinh2014normalizingflows} that are each parametrized by monotonic rational quadratic splines \cite{NEURIPS20197ac71d43}. As a reminder, $\gvb$ applies a coordinate transformation on $\zvb\in \mathbb{R}^D$. 

\begin{figure}
    \centering
    \includegraphics[width = 0.46\textwidth]{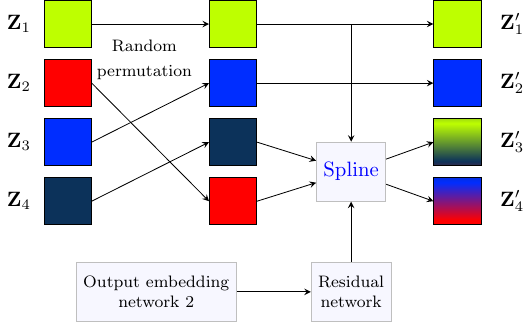}
    \caption{Schematic overview of an elementary cell of the coupling transform on the example of a four-dimensional flow. The compressed data that is summarized by the second embedding network serves as the input data for the spline through a residual neural network. }
    \label{ml: app: fig: schematic plot elementary cell coupling flows}
\end{figure}

One can write the sequence of coupling transforms as 
\begin{equation}
    \gvb = 
    \gvb_{(\nblock)} \circ \gvb_{(\nblock - 1)} 
    \circ
    \ldots
    \circ
    \gvb_{(2)} \circ \gvb_{(1)}\,,
\end{equation}
where $\nblock$ is the \textit{number of blocks}. Each of the functions $\gvb_{(i)}$ depends on the data, a set of \gls{nn} parameters and the latent variable $\zvb$. 
The $\gvb_i$ all share the same functional form (but have different model parameters). First, one applies a random (but different for each coupling transform, fixed during training) permutation of $\zvb$. Then the first $k$ components of the reshuffled variable $\tilde \zvb$ are left unchanged. The remaining parameters undergo an invertible (spline) transformation $\svb_i$ that is parametrized by the \gls{nn} parameters $\Theta_i$.
One can write the transformation component-wise as \cite{dinh2014normalizingflows}
\begin{equation}
    \left[\gvb_{(i)}(\zvb)\right]_j = 
    \begin{cases}
        \zvb_j
        \quad &\text{if }1\leq j \leq k \,,
        \\
        \left[
        \svb(\zvb;\Theta_{(i)}) 
        \right]_j
        \quad &\text{if }k+1\leq j \leq D \,,
    \end{cases}
\end{equation}
where $\Theta_{(i)}$ are the \gls{nn} parameters of the $i$th coupling transformation. 
A fully-connected residual network governs the spline function. 
Each transformation is invertible, and its inverse and Jacobian is simple to compute. 
Fig.~\ref{ml: app: fig: schematic plot elementary cell coupling flows} shows schematically the transformation of one such elementary cells that make up the flow. 
In our case, the transformation $\svb:\mathbb{R}^{D}\rightarrow \mathbb{R}^{D-j}$ corresponds to a monotonic\footnote{If the spline was not monotonic the function would be not invertible, making it unsuitable for \glspl{nf}.}, quadratic, rational spline function \cite{NEURIPS20197ac71d43} (see Fig.~1 of \cite{NEURIPS20197ac71d43} for an example of a one-dimensional spline function). The parameters of this function are governed by a \gls{nn}. Note that $\svb$ depends on all variables $\zvb$ and can hence incorporate correlations between different parameters. 

\section{Loss identity}
\label{app: methods: machine learning: loss lemma}

In this appendix, we show that the loss function proposed in Eq.\,\eqref{methods: machine learning: eq: theoretical loss for model} equals the expectation value of the \gls{kl} divergence between the true posterior and the model. 
We proceed in two steps: we demonstrate that the four expectation values can be rewritten in terms of two expectation values. These two expectation values can then be exchanged (from Bayes' theorem), yielding the desired result.

Let us consider the following expectation value of the function $f(w,z)$ 
\begin{equation}
    \loss=\mathbb{E}_{p(w)}
  \mathbb{E}_{p(x|w)}\mathbb{E}_{p(y|x)}\mathbb{E}_{p(z|y)}f(w,z)\,.
\end{equation}
This can be written from definition as a fourfold integration
\begin{equation}
    \loss=\int \mathrm{d}w \,
    \mathrm{d}x\,
    \mathrm{d}y\,
    \mathrm{d}z\,
    p(w)\,
p(x|w)\,p(y|x)\,p(z|y)\,f(w,z)\,.
\end{equation}
Making the additional assumption that $y$ and $w$ are conditionally independent given $x$, i.e.,
\begin{equation}
p(y, w | x) = p(y | x) p(w | x) \qquad \Leftrightarrow \qquad p(y | x,w) = p(y | x),
\end{equation}
and similarly that $z$ and $(x, w)$ are conditionally independent given $y$, so that $p(z|y) = p(z|y,x,w)$, and applying the law of  conditional probability $p(s|t)p(t)=p(s,t)$, the above expression can be rewritten as 
\begin{equation}
    \loss=\int \mathrm{d}w \,
    \mathrm{d}x\,
    \mathrm{d}y\,
    \mathrm{d}z\,
    p(w,x,y,z)\,f(w,z)\,.
\end{equation}
Since the function $f$ is independent of the random variables $x,y$, we can perform the integration
\begin{equation}
    \loss=\int \mathrm{d}w \,
    \mathrm{d}z\,
    p(w)\,\,p(z|w)\,f(w,z) 
\end{equation}
After applying Bayes' theorem and changing the order of the integration, we obtain
\begin{equation}
    \loss=\int \mathrm{d}w\,\mathrm{d}z \,p(z)\,p(w|z)f(w,z)=\mathbb{E}_{p(z)}
    \mathbb{E}_{p(w|z)}f(w,z)\,.
\end{equation}
The identity claimed in Section~\ref{subsec: divide and conquer} of equations~\eqref{methods: machine learning: eq: theoretical loss for model} and \eqref{methods: machine learning: eq: loss rewritten 2 expectation values} can be obtained for $w=\Lambda$, $x=\theta_K$, $y = \data_K$, $z=\thetapsK$, $f = -\log[q(\Lambda|\thetapsK)]$ and $p(w|z)$ is the target distribution, namely the population posterior given a set of GW events $p(\Lambda|\setdata)$. The conditional independence conditions reduce to assuming $p(\thetapsK, \Lambda, \theta_K | \data_K) = p(\thetapsK | \data_K) p(\Lambda, \theta_K | \data_K) $, which holds because the distribution of posterior samples depends only on the observed data, and $p(\data_K, \Lambda | \theta_K ) = p(\data_K | \theta_K) p(\Lambda | \theta_K)$, which holds because the observed data depends only on the parameters of the sources in the data.

\section{Population details}
\label{app: summary populations}

In the results section we analyze 13 different populations, each of which has a different underlying set of \metaps{}. Table~\ref{tab: summary populations} lists the value of all parameters for each population.

\begin{table}[]
    \centering
    \begin{tabular}{cccccc}
        \hline
        \hline
        $\Lambda$ &  $H_0$ &  $m_{\rm min}$ &  $m_{\rm max}$ &  $\alpha$ &  $\beta$ \\
        \hline
\diagbox{\# Pop}{Units} & {\specialcell{${\rm km\,s^{-1}}\times$ \\ $ {\rm Mpc}^{-1}$}} & $\msun$ & $\msun$ & - & - \\
        \hline
            0 & 60.0 & 22.0 & 46.0 & 0.0 & 0.0 \\
1 & 50.0 & 21.0 & 44.0 & 0.5 & 0.5 \\
2 & 55.0 & 23.0 & 42.0 & 1.0 & -0.5 \\
3 & 100.0 & 18.5 & 46.0 & 1.0 & 0.0 \\
4 & 50.0 & 24.0 & 45.0 & 1.0 & 0.2 \\
5 & 90.0 & 23.0 & 44.0 & 0.4 & 0.0 \\
6 & 80.0 & 24.0 & 43.0 & -0.3 & -0.7 \\
7 & 45.0 & 19.0 & 38.0 & 1.5 & 0.5 \\
8 & 135.0 & 19.0 & 46.0 & 1.5 & -0.5 \\
9 & 60.0 & 19.0 & 45.0 & -1.6 & 0.0 \\
10 & 70.0 & 22.0 & 43.0 & -0.4 & 0.4 \\
11 & 80.0 & 19.4 & 42.3 & 0.6 & -0.3 \\
12 & 67.0 & 20.1 & 42.9 & 0.6 & -0.5 \\
13 & 70.0 & 21.0 & 43.3 & 0.8 & 0.3 \\
        \hline
        \hline
    \end{tabular}
    \caption{The \metaps{} for the fourteen populations considered. }
    \label{tab: summary populations}
\end{table}

\bibliography{references}

\end{document}